\documentclass[useAMS,usenatbib]{mnras}

\usepackage{ulem}
\usepackage{color}
\usepackage{graphics,graphicx}
\usepackage{times} 
\usepackage{amssymb}
\usepackage{amsmath}
\usepackage{natbib}
\usepackage{url}
\usepackage{multirow}
\usepackage{ctable}
\usepackage{threeparttable}
\newif\ifAMStwofonts
\AMStwofontstrue

\def\asca{{\it ASCA}}

\def\xmm{{\it XMM-Newton}}

\def\suzaku{{\it Suzaku}}

\def\bepposax{{\it BeppoSAX}}

\def\epicpn{{EPIC-pn}}
\def\epicmos1{{EPIC-MOS1}}
\def\epicmos2{{EPIC-MOS2}}
\def\epicmos{{EPIC-MOS}}

\def\pin{{\rm PIN}}
\def\hxd{{\rm HXD}}
\def\nustar{{\it NuSTAR}}

\def\deg{$^{\circ}$}

\def\kmps{\hbox{$\rm\thinspace km~s^{-1}$}}
\def\pcmsq{\hbox{$\rm\thinspace cm^{-2}$}}
\def\H0{{\rm ~km~s^{-1}~Mpc^{-1}}}

\def\kev{\hbox{\rm keV}}

\def\ctps{\hbox{$\rm\thinspace ct~s^{-1}$}}

\def\ergpcmsqps{\hbox{$\rm\thinspace erg~cm^{-2}~s^{-1}$}}

\def\ergps{\hbox{erg~s$^{-1}$}}
\def\ergcmps{\hbox{\rm erg~cm~s$^{-1}$}}

\def\msun{\hbox{$\rm M_{\odot}$}}

\def\chisq{{$\chi^{2}$}}

\def\xspecv{\hbox{\small XSPEC}\, v12.6.0f}

\def\heasoft{\hbox{\rm{\small HEASOFT}}}
\def\nustardas{\rm {\small NUSTARDAS}}
\def\xselect{\hbox{\rm{\small XSELECT}}}
\def\xmmselect{\hbox{\rm{\small XMMSELECT}}}
\def\ftool{\hbox{\rm{\small FTOOL}}}

\def\addascaspec{\hbox{\rm{\small ADDASCASPEC~\/}}}
\def\flx2xsp{\rm{\small FLX2XSP}}
\def\sas{\hbox{\rm{\small SAS~\/}}}
\def\epchain{\hbox{\rm{\small EPCHAIN}}}
\def\emchain{\hbox{\rm{\small EMCHAIN}}}

\def\rmfgen{\hbox{\rm{\small RMFGEN}}}
\def\arfgen{\hbox{\rm{\small ARFGEN}}}

\def\addascaspec{\hbox{\rm{\small ADDASCASPEC}}}
\def\nupipeline{\rm{\small NUPIPELINE}}
\def\nuproducts{\rm{\small NUPRODUCTS}}

\def\xstar{\hbox{\rm{\small XSTAR}}}
\def\grid25{\hbox{\rm{\small GRID25}}}

\def\cabs{\rm{\small CABS}}

\def\tbnew{\rm{\small TBNEW}}
\def\tbnewpcf{\rm{\small TBNEW\_PCF}}

\def\xillver{\rm{\small XILLVER}}
\def\xillvercp{\rm{\small XILLVER\_CP}}

\def\relconv{\rm{\small RELCONV}}

\def\relxill{\rm{\small RELXILL}}
\def\relxilllp{{\small RELXILLLP}}
\def\relxilllpcp{\rm{\small RELXILLLP\_CP}}

\def\mytorus{\rm{\small MYTORUS}}

\def\nthcomp{\rm{\small NTHCOMP}}

\def\partcov{\rm{\small PARTCOV}}

\def\civ{\hbox{\rm C\,{\small IV}}}
\def\fexxv{\hbox{\rm Fe\,{\small XXV}}}
\def\fexxvi{\hbox{\rm Fe\,{\small XXVI}}}

\def\heii{\rm He\,{\small II}}
\def\ciii{\rm C\,{\small III}}
\def\civ{\rm C\,{\small IV}}

\def\eg{{\it e.g.}}
\def\etc{{\it etc.}}
\def\ie{{\it i.e.~\/}}

\def\la{\mathrel{\hbox{\rlap{\hbox{\lower4pt\hbox{$\sim$}}}{\raise2pt\hbox{$<$}}}}}
\def\ga{\mathrel{\hbox{\rlap{\hbox{\lower4pt\hbox{$\sim$}}}{\raise2pt\hbox{$>$}}}}}

\def\d25{D$_{25}$}
\def\nh{{$N_{\rm H}$}}

\def\.25{0.25 keV\thinspace}

\def\rg{$R_{\rm{G}}$}
\def\rh{$R_{\rm{H}}$}

\def\Rfrac{$R_{\rm{frac}}$}

\def\iras{IRAS\,13197-1627}

\title[Broadband Spectroscopy of IRAS\,13197-1627]{Disentangling the Complex Broadband X-ray Spectrum of IRAS\,13197-1627 with \textit{NuSTAR}, \textit{XMM-Newton} and \textit{Suzaku}}

\author[D.\,J. Walton et al.]
{\parbox{7.in}{D.\,J. Walton$^{1}$ \thanks{E-mail: dwalton@ast.cam.ac.uk},
M. Brightman$^{2}$,
G. Risaliti$^{3}$,
A. C. Fabian$^{1}$,
F. F\"urst$^{4}$,
F. A. Harrison$^{2}$,
A. Lohfink$^{1}$,
G. Matt$^{5}$,
G. Miniutti$^{6}$,
M. L. Parker$^{1}$,
D. Stern$^{7}$ \\
\\[-0.2cm]
\footnotesize
$^{1}$ \it{Institute of Astronomy, University of Cambridge, Madingley Road, Cambridge CB3 0HA, UK} \\
$^{2}$ \it{Space Radiation Laboratory, California Institute of Technology, Pasadena, CA 91125, USA} \\
$^{3}$ \it{Dipartimento di Fisica e Astronomia, Universita di Firenze, via G. Sansone 1, 50019 Sesto Fiorentino, Firenze, Italy} \\
$^{4}$ \it{European Space Astronomy Centre (ESA/ESAC), Operations Department, Villanueva de la Ca\~nada (Madrid), Spain} \\
$^{5}$ \it{Dipartimento di Matematica e Fisica, Universita degli Stu di Roma Tre, via della Vasca Navale 84, 00146 Roma, Italy} \\
$^{6}$ \it{Centro de Astrobiolog\'{i}a (CSIC--INTA), Dep. de Astrof\'{i}sica, ESAC campus, Camino Bajo del Castillo s/n, E-28692 Villanueva de la Ca\~nada, Spain} \\
$^{7}$ \it{Jet Propulsion Laboratory, California Institute of Technology, Pasadena, CA 91109, USA}}}

\date{}

\begin{document}
\pagerange{\pageref{firstpage}--\pageref{lastpage}}
\maketitle
\label{firstpage}

\begin{abstract}
We present results from a coordinated \xmm+\nustar\ observation of the type 1.8
Seyfert galaxy \iras. This is a highly complex source, with strong contributions
from relativistic reflection from the inner accretion disk, neutral absorption and
further reprocessing by more distant material, and ionised absorption from an
outflow. We undertake a detailed spectral analysis combining the broadband
coverage provided by \xmm+\nustar\ with a multi-epoch approach incorporating
archival observations performed by \xmm\ and \suzaku. Our focus is on
characterising the reflection from the inner accretion disk, which previous works
have suggested may dominate the AGN emission, and constraining the black hole
spin. Using lamppost disk reflection models, we find that the results for the inner
disk are largely insensitive to assumptions regarding the geometry of the distant
reprocessor and the precise form of the illuminating X-ray continuum. However,
these results do depend on the treatment of the iron abundance of the distant
absorber/reprocessor. The multi-epoch data favour a scenario in which the AGN
is chemically homogeneous, and we find that a rapidly rotating black hole is
preferred, with $a^* \geq 0.7$, but a slowly-rotating black hole is not strongly
excluded. In addition to the results for the inner disk, we also find that both the
neutral and ionised absorbers vary from epoch to epoch, implying that both have
some degree of inhomogeneity in their structure.
\end{abstract}


\section{Introduction}

Relativistically broadened iron emission is often seen from active galactic nuclei
(AGN), both from individual sources (\eg\ \citealt{Tanaka95, FabZog09, Reis14nat,
Reynolds14, Xu17iras}) and stacked spectra from AGN samples (\eg\
\citealt{Walton15lqso, Mantovani16}). Most AGN also show a `hard' excess above
10\,\kev\ consistent with Compton reflection from the accretion disk (\eg
\citealt{Walton10Hex, Nardini11, Rivers13}). Together, these disk reflection features
offer an opportunity to constrain the spins of the supermassive black holes that
power these sources (\eg\ \citealt{Fabian89, kdblur, Dauser14}), providing a rare
observational window into their formation history (\eg\ \citealt{Volonteri13, Sesana14,
Dubois14}). To date, spin estimates for $\sim$30 AGN have been obtained through
study of disk reflection, suggesting a preference for rapidly rotating black holes (\eg\
\citealt{Walton13spin, Reynolds14rev, Vasudevan16}). However, the sample is
relatively small and still not well defined in a statistical sense. Strong selection biases
likely exist, as higher spin sources are expected to be brighter for a given rate of
accretion (\citealt{Brenneman11, Reynolds12}), and it is also easier to obtain tight
constraints on the spin parameter for more rapidly rotating black holes
(\citealt{Walton13spin, Bonson16}). It is therefore vital to further expand the sample
of spin measurements such that these biases may be overcome.

\begin{table*}
  \caption{Details of the X-ray observations of IRAS\,13197-1627 considered
in this work.}
%\vspace{-0.6cm}
\begin{center}
\begin{tabular}{c c c c c c}
\hline
\hline
\\[-0.1cm]
Epoch & Mission & OBSID & Start & Exposure\tmark[a] & Count Rate\tmark[b] \\
& & & Date & (ks) & (\ctps) \\
\\[-0.2cm]
\hline
\hline
\\[-0.2cm]
\multicolumn{5}{c}{\textit{2016}} \\
\\[-0.3cm]
\multirow{2}{*}{XN1} & \nustar\ & 60101020002 & \multirow{2}{*}{2016-01-17} & 85 & $0.121 \pm 0.001$ \\
\\[-0.3cm]
& \xmm\ & 0763220201 & & 107/132 & $0.118 \pm 0.001$ \\
\\[-0.2cm]
\multicolumn{5}{c}{\textit{Archival}} \\
\\[-0.3cm]
X1 & \xmm\ & 0206580101 & 2005-01-24 & 32/39 & $0.166 \pm 0.002$ \\
\\[-0.3cm]
X2 & \xmm\ & 0506340101 & 2008-01-24 & 57/76 & $0.179 \pm 0.002$ \\
\\[-0.3cm]
S1 & \suzaku\ & 704022010 & 2009-07-01 & 42/34 & $0.045 \pm 0.001$ \\
\\[-0.3cm]
S2 & \suzaku\ & 707027010 & 2013-01-04 & 156/127 & $0.0184 \pm 0.0004$ \\
\\[-0.3cm]
\hline
\hline
%\\[0.4cm]
\end{tabular}
\end{center}
\flushleft
$^{a}$ \xmm\ exposures are listed for the \epicpn/MOS detectors, and
\suzaku\ exposures are given for the XIS/PIN detectors. \\
$^{b}$ Observed count rates are given in the 3--10\,keV band for \xmm\ (\epicpn) and
\suzaku (XIS0), and the 3.5--79\,keV band for \nustar\ (FPMA).
\vspace*{0.3cm}
\label{tab_obs}
\end{table*}

In part, the sample size is hindered because the detection and characterisation
of emission from the inner disk can be complicated by both absorption from
material along our line-of-sight to the inner accretion flow and additional
reprocessing by more distant structures, which can also introduce strong spectral
features in the X-ray band (\eg\ \citealt{Miller08, LMiller09, Sim10}). Robustly
disentangling the absorption and reflection requires both sensitive coverage
across all the key reflection features and good spectral resolution. As such, the
combination of \xmm\ and \nustar\ provides a particularly powerful tool for the study
of AGN. The continuous 3--79\,keV bandpass of \nustar\ (\citealt{NUSTAR}) is well
suited to the study of reflection, while the better spectral resolution of \xmm\
(\citealt{XMM}) helps to separate out the broadened iron emission from the
signatures of absorption. Even in cases that exhibit complex absorption, as is seen
for example in the well-studied Seyfert galaxy NGC\,1365 (\eg\ \citealt{Rivers15}),
coordinated \xmm\ and \nustar\ observations can often isolate the relativistic
reflection from the inner disk (\eg\ \citealt{Risaliti13nat, Walton14}).

\iras\ (also known as MCG--03-34-64) is a source of particular interest in this
respect. \iras\ is a bright, nearby ($z=0.01654$), type 1.8 Seyfert AGN with a
complex X-ray spectrum. Previous observations with \xmm, \suzaku, \bepposax\
and \asca\ have shown that \iras\ is typically absorbed by a moderate neutral
column ($N_{\rm{H}} \sim 5 \times 10^{23}$ cm$^{-2}$, although on rare
occasions it can exhibit a much greater level of obscuration, reaching column
densities up to $N_{\rm{H}} \sim 10^{24}$ cm$^{-2}$), and that it also exhibits
evidence for a relativistically broadened iron line from the innermost accretion
flow (\citealt{Dadina04, Miniutti07iras, torus}). In addition to the broad iron line,
there is evidence for narrow absorption features from blueshifted ionised iron at
$\sim$7--7.5\,keV, and a clear narrow core to the iron emission. In many respects,
\iras\ therefore shows many similarities to the NGC\,1365. Most importantly, the
high-energy data from the non-imaging \suzaku\ PIN and \bepposax\ PDS
detectors show evidence for an extremely strong excess over the expected
powerlaw-like AGN continuum (e.g. \citealt{Tatum13}), potentially suggesting an
unusually strong contribution from reflection (\citealt{Miniutti07iras}). If this is
correct, the implied accretion geometry is rather extreme, requiring a compact
X-ray source located very close to the black hole such that gravitational
lightbending can suppress the primary continuum emission relative to the
reflection from the inner disk (\eg\ \citealt{lightbending}). However, \iras\ is located
$\sim$2$'$ away from another active galaxy, MCG--3-34-63, potentially resulting
in source confusion issues with the older non-imaging hard X-ray PIN and PDS
detectors. Hard X-ray imaging spectrometers are required to unambiguously
confirm the flux observed by these detectors is associated with \iras.

Here we present results from a coordinated broadband observation of \iras\ with
both \nustar\ and \xmm\ in order to disentangle the complex X-ray spectrum
exhibited by this source and test the extreme reflection scenario. The rest of the
paper is structured as follows: in section \ref{sec_red} we describe the \xmm\ and
\nustar\ observations and outline our data reduction procedure, in sections
\ref{sec_image} and \ref{sec_spec}  we present our analysis of these data, and in
section \ref{sec_dis} we discuss the results obtained and summarise our
conclusions.

\begin{figure*}
\begin{center}
%\hspace*{-0.5cm}
\rotatebox{0}{
{\includegraphics[width=235pt]{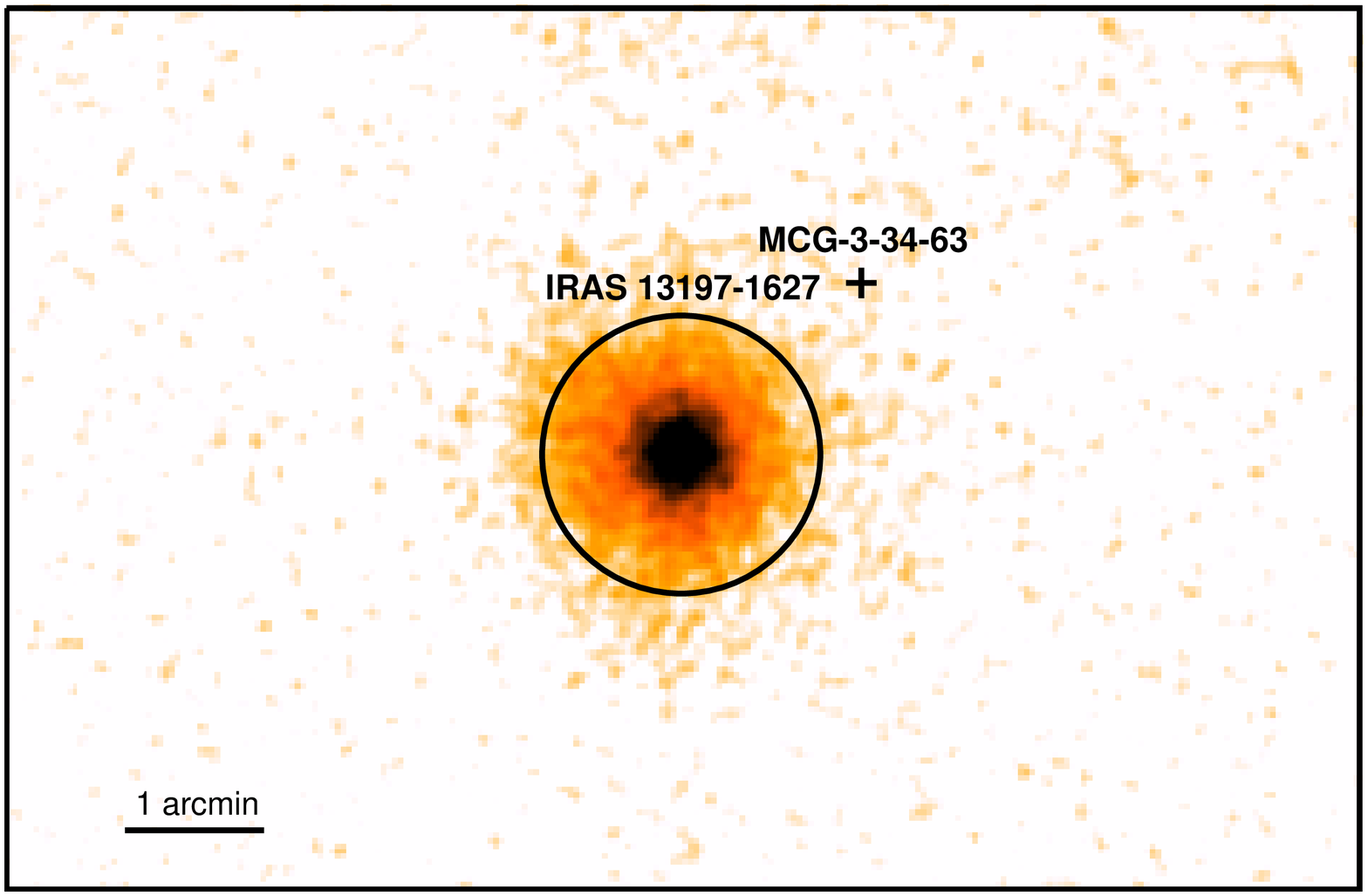}}
}
\hspace{0.35cm}
\rotatebox{0}{
{\includegraphics[width=235pt]{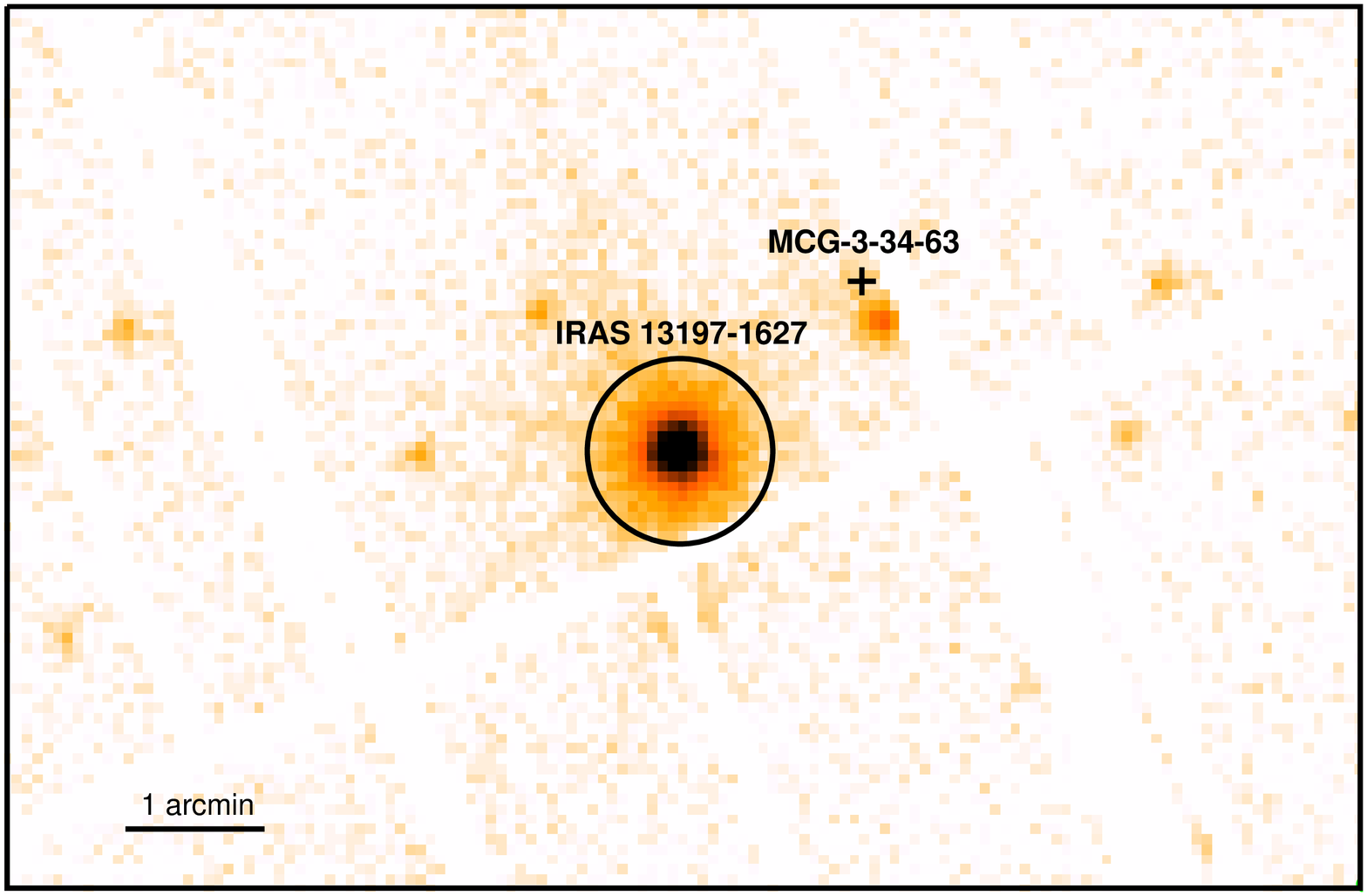}}
}
\end{center}
\caption{
\nustar\ (\textit{left}, FPMA, 3--79\,keV) and \xmm\ (\textit{right}, \epicpn,
0.3--10\,keV) images of \iras. The position of the nearby AGN MCG--3-34-63, which is
not detected by \nustar, is also shown. The hard X-ray flux from this field is dominated
by \iras. The circular regions indicate the source extraction regions used in each case.}
\label{fig_image}
\end{figure*}

%\begin{figure}
%%\hspace*{-0.5cm}
%\epsscale{1.05}
%\plotone{./figs/iras13197_nustar_image.eps}
%\vspace*{0.15cm}
%\caption{\nustar\ image of \iras. The position of the nearby AGN MCG--3-34-63,
%which is not detected, is also shown. The hard X-ray flux from this field is
%dominated by \iras.
%}
%\vspace{0.3cm}
%\label{fig_image}
%\end{figure}

\section{Observations and Data Reduction}
\label{sec_red}

\subsection{2016 Observation}

\iras\ was observed simultaneously with \nustar\ and \xmm\ and on 2016 January 17;
see Table \ref{tab_obs} for details. The following sections detail our reduction of the
data from this coordinated observation, referred to hereafter as epoch XN1.

\subsubsection{NuSTAR}

We reduced the \nustar\ data following standard procedures, first cleaning the data
with \nupipeline, part of the \nustar\ Data Analysis Software (\nustardas, v1.6.0). We
used the standard depth correction, which significantly reduces the internal
high-energy background, and also removed passages through the South Atlantic
Anomaly. Instrumental calibration files from \nustar\ caldb v20160824 are used
throughout this work. Source and background spectra/lightcurves and instrumental
responses were then produced for each of the two focal plane modules (FPMA/B)
using \nuproducts. Source products were obtained from circular regions of radius
60$''$, and background was estimated from larger regions of blank sky on the same
detector as \iras. In order to maximise the signal-to-noise (S/N), we also extracted
the `spacecraft science' (mode 6) data in addition to the standard `science' (mode
1) data, following the method outlined in \cite{Walton16cyg}, which in this case
provides $\sim$10\% of the total $\sim$85\,ks good \nustar\ exposure. \iras\ is
detected across the entire \nustar\ band (the S/N above 30\,keV is $\sim$20 for
each FPM), and we analyse the \nustar\ data between 3.5--79\,keV owing to
a slight ($\sim$40\%) discrepancy between \nustar\ and \xmm\ below 3.5\,keV in
FPMB. This is likely related to the fact that the point-spread function (PSF) is
energy-dependent at very low energies (\citealt{NUSTARcal}), and on FPMB the
PSF for \iras\ straddles a chip gap, resulting in an uncertain PSF correction below
$\sim$3.5\,keV in this case.

\subsubsection{XMM-Newton}
\label{sec_xmmred}

The \xmm\ data reduction was carried out with the \xmm\ Science Analysis
System (\sas v15.0.0), following the standard prescription provided in the online
guide.\footnote{https://www.cosmos.esa.int/web/xmm-newton} Raw data files were
cleaned using \epchain\ for the \epicpn\ detector (\citealt{XMM_PN}), and \emchain\
for two \epicmos\ units (\citealt{XMM_MOS}). Source products were extracted from
the cleaned event files from circular regions of radius 40$''$ and 45$''$ for \epicpn\
and \epicmos, respectively, using \xmmselect. As with the \nustar\ data,
background was estimated from larger regions of blank sky on the same chip as
\iras. Only single and double events were considered for \epicpn\ and single to
quadruple events were considered for \epicmos. Periods of high background
were excluded as standard. Instrumental response files for each of the detectors
were generated with \rmfgen\ and \arfgen. After performing the reduction
separately for the two \epicmos\ detectors, and confirming their consistency,
these data were combined into a single \epicmos\ spectrum using \addascaspec.

\subsection{Archival Data}

In addition to the new 2016 observation, we also analyse several archival
observations of \iras, focusing on high signal-to-noise data obtained with the more
recent generation of X-ray observatories. \iras\ was observed twice with \xmm\
alone (epochs X1, X2), and also twice by \suzaku\ (epochs S1, S2; see Table
\ref{tab_obs}). Epochs X1 and S2 have previously been studied by
\cite{Miniutti07iras} and \cite{torus}, and by \cite{Tatum13}, respectively. The
following sections provide details on our reduction of these data.

\subsubsection{XMM-Newton}

Our reduction procedure for the archival \xmm\ observations largely follows that of
the 2016 data, outlined in Section \ref{sec_xmmred}. For OBSID 0206580101 (epoch
X1), we used the same source extraction regions as for the 2016 data. However, for
OBSID 0506340101 (epoch X2) we used slightly smaller source regions of radius
30$''$ for the \epicpn\ detector to avoid chip gaps, as \iras\ was placed slightly off-axis.

\subsubsection{Suzaku}

The data from the two \suzaku\ observations were reduced using the \heasoft\
software package (v6.19), following the procedure outlined in the \suzaku\ Data
Reduction Guide\footnote{http://heasarc.gsfc.nasa.gov/docs/suzaku/analysis/}. To
extract science products from the XIS units (\citealt{SUZAKU_XIS}), we reprocessed
the unfiltered event files for each of the operational CCDs (XIS0, 1, 3) and editing
modes (3x3, 5x5). Cleaned event files were generated by running the \suzaku\
pipeline with the latest calibration and screening criteria files (XIS caldb v20160607).
Source products were extracted from circular regions $\sim$150$''$ in radius using
\xselect, and the background was extracted from adjacent regions free of any
contaminating sources, with care taken to avoid the calibration sources in the
corners. Instrumental response files were generated for each detector using the
{\small XISRESP} script with a medium resolution. The spectra and response files
for the two front-illuminated (FI) detectors (XIS0, 3) were combined using the \ftool\
\addascaspec; the XIS1 unit is the only back-illuminated (BI) detector.

For the \hxd\ \pin\ detector (\citealt{SUZAKU_HXD}) we also reprocessed the
unfiltered event files with the latest calibration/screening files (HXD caldb v20110913).
The \hxd\ is a collimating instrument, so background estimation requires separate 
consideration of both the non X-ray instrumental background (NXB) and the cosmic
X-ray background (CXB). The instrumental response and NXB background files are 
provided by the \suzaku\ team for each
observation\footnote{http://www.astro.isas.ac.jp/suzaku/analysis/hxd/}; in this work
we use the higher quality `tuned' (Model D) background model. The final spectral
products were generated using the {\small HXDPINXBPI} script which, in addition to
extracting the source spectrum, adds a simulated contribution from the CXB to the
NXB (using the form of \citealt{Boldt87}) to produce a total background spectrum.
We analyse the \pin\ data over the $\sim$15--50\,keV bandpass.

\section{\textit{NuSTAR} Imaging}
\label{sec_image}

As noted previously, \iras\ is located $\sim$2$'$ away from MCG--3-34-63, another
known X-ray emitting active galaxy ($z = 0.02133$, classified as a Seyfert 2). While
the latter AGN is significantly fainter than \iras\ in the soft X-ray band (typically by
more than a factor of 100), sources with extremely hard spectra that are only really
seen in hard X-rays are known (\eg\ \citealt{Lansbury17}), and the only previous hard
X-ray ($>$10\,keV) observations of this field have been with non-imaging instruments
(the PDS detector on \bepposax\ and the PIN detector on board \suzaku:
\citealt{Miniutti07iras, Tatum13}), leaving a major uncertainty over their relative
contributions in the band in which the hard excess is seen. \nustar\ is the first hard
X-ray mission with sufficient imaging capabilities to resolve \iras\ and MCG--3-34-63.
We show the \nustar\ image in Figure \ref{fig_image}. It is clear that the hard X-ray
flux is entirely dominated by \iras\ (MCG--3-34-63 is not even detected), confirming
the hard X-ray emission recorded by \bepposax\ and \suzaku\ is indeed associated
with \iras. Assuming that \iras\ constantly outshines MCG--3-34-63 by a significant
factor at these energies, we adopt the standard cross-calibration constants of 1.18 for
epoch S1 (performed with an XIS nominal pointing) and 1.16 for epoch S2 (performed
with an HXD nominal pointing) for the \suzaku\ PIN detector with respect to the XIS
units during our analysis of the archival data (Section \ref{sec_oldobs})

\section{Spectral Analysis}
\label{sec_spec}

The majority of our work focuses on spectral analysis of the observations considered.
Model fits are performed with \xspecv\ (\citealt{xspec}), and we quote parameter
uncertainties at the 90\% confidence level for one interesting parameter. All models
include a Galactic absorption component with a fixed column of $N_{\rm{H,Gal}} = 4.99
\times 10^{20}$\,\pcmsq\ (\citealt{NH}), modelled with the \tbnew\ neutral absorption
code (\citealt{tbabs}). We use the cross-sections of \cite{Verner96} for the absorption,
as recommended, but we combine these with the solar abundance set of
\cite{Grevesse98} for internal self-consistency with both the \xillver\ family of reflection
models (\citealt{xillver}) and the \xstar\ photoionisation code (\citealt{xstar}), which are
heavily utilized throughout this work.

\subsection{2016 Data}
\label{sec_XN1}

\begin{figure}
\begin{center}
\hspace*{-0.4cm}
\rotatebox{0}{
{\includegraphics[width=235pt]{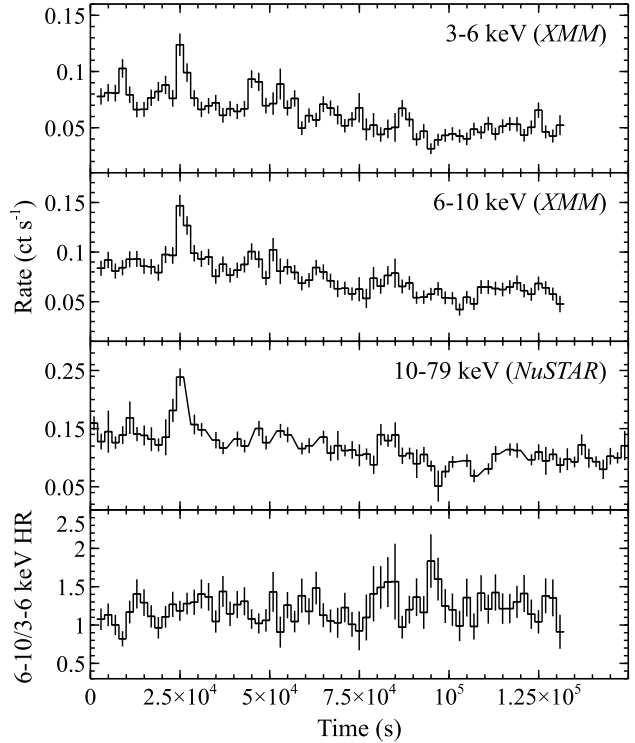}}
}
\end{center}
\caption{
The lightcurves observed with the \xmm\ \epicpn\ detector in the 3--6 and 6--10\,keV
bands, and with the \nustar\ FPMA detector in the 10-79 keV band (2\,ks time bins;
first three panels). The bottom panel shows the hardness ratio computed between the
6--10 and 3--6\,keV \xmm\ bands. Although some flux variability is observed, including
a small flare towards the beginning of the observation, no evidence for strong spectral
variability is seen.}
\label{fig_lc}
\end{figure}

We begin our analysis by focusing on the new coordinated observation taken with
\xmm+\nustar\ (XN1). In Figure \ref{fig_lc} we show \xmm\ (EPIC-pn) and \nustar\
(FPMA) lightcurves from this observation extracted in the 3--6, 6--10 and 10--79\,keV
bands, as well as a simple hardness ratio computed as the ratio between the EPIC-pn
count rates in the 6--10\,keV and 3--6\,keV bands. While there is some flux variability
observed, including a short and fairly sharp flare towards the beginning of the
observation which seen by both missions, there is little evidence for strong spectral
variability, so we focus on modelling the time-averaged spectrum from this epoch. In
our analysis of these data, cross-calibration uncertainties between the different
detectors are accounted for by allowing multiplicative constants to float between the
datasets, fixing \epicpn\ at unity. These constants are within $\sim$10\% of unity, as
expected (\citealt{NUSTARcal}).

%\begin{figure}
%\hspace*{-0.5cm}
%\epsscale{1.05}
%\plotone{./figs/iras13197_xmm_HR_lc_2ks.eps}
%\caption{\textit{Top panel:} the lightcurve observed with the \epicpn\ detector in
%the 6--10\,keV bandpass (2\,ks time bins). \textit{Bottom panel:} hardness ratio
%computed between the 6--10 and 3--6\,keV bands. Although some flux variability
%is observed, including a small flare towards the beginning of the observation, no
%evidence for strong spectral variability is seen.}
%\vspace{0.3cm}
%\label{fig_lc}
%\end{figure}

\begin{figure*}
\begin{center}
\hspace*{-0.3cm}
\rotatebox{0}{
{\includegraphics[width=235pt]{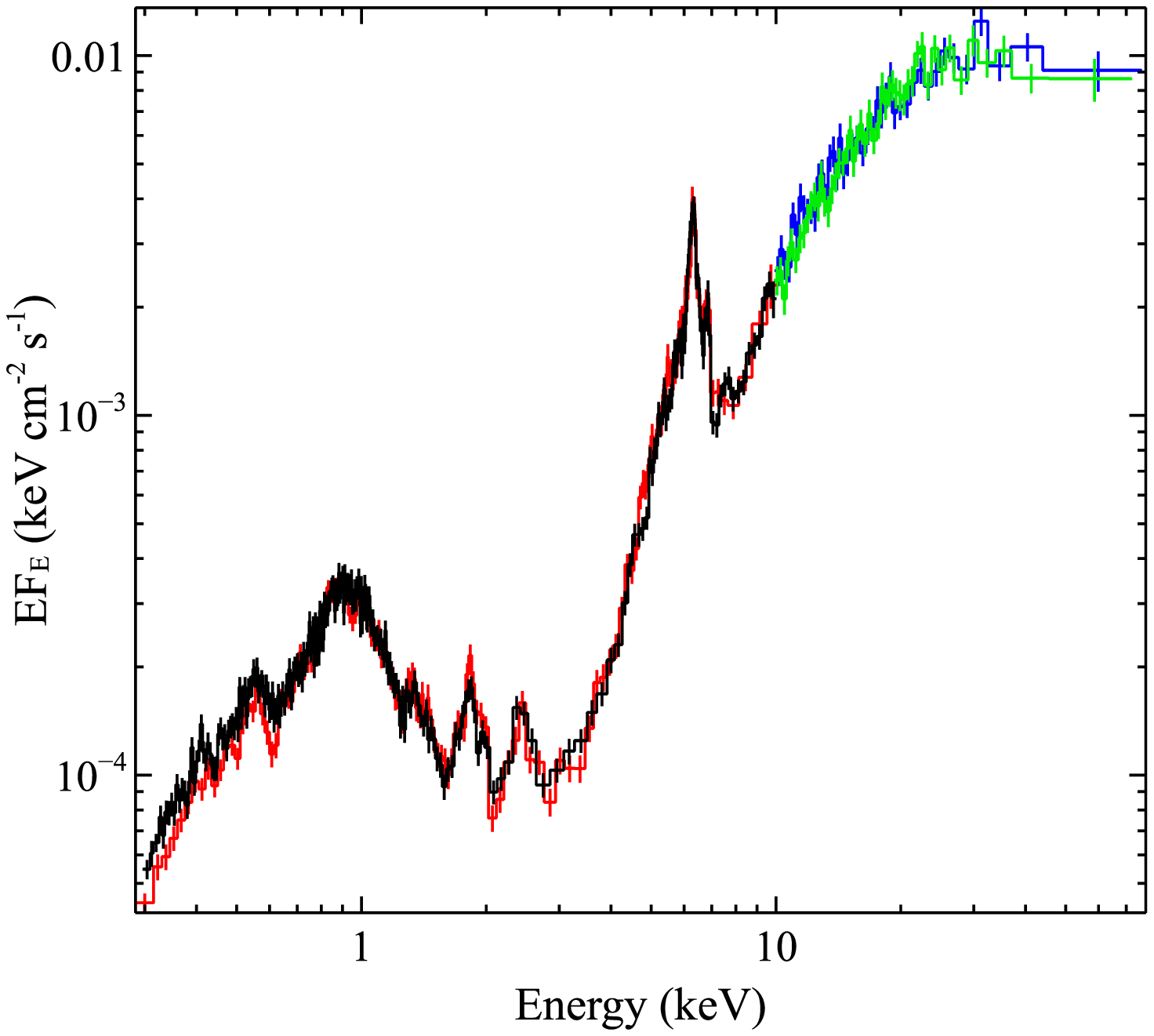}}
}
\hspace*{0.5cm}
\rotatebox{0}{
{\includegraphics[width=235pt]{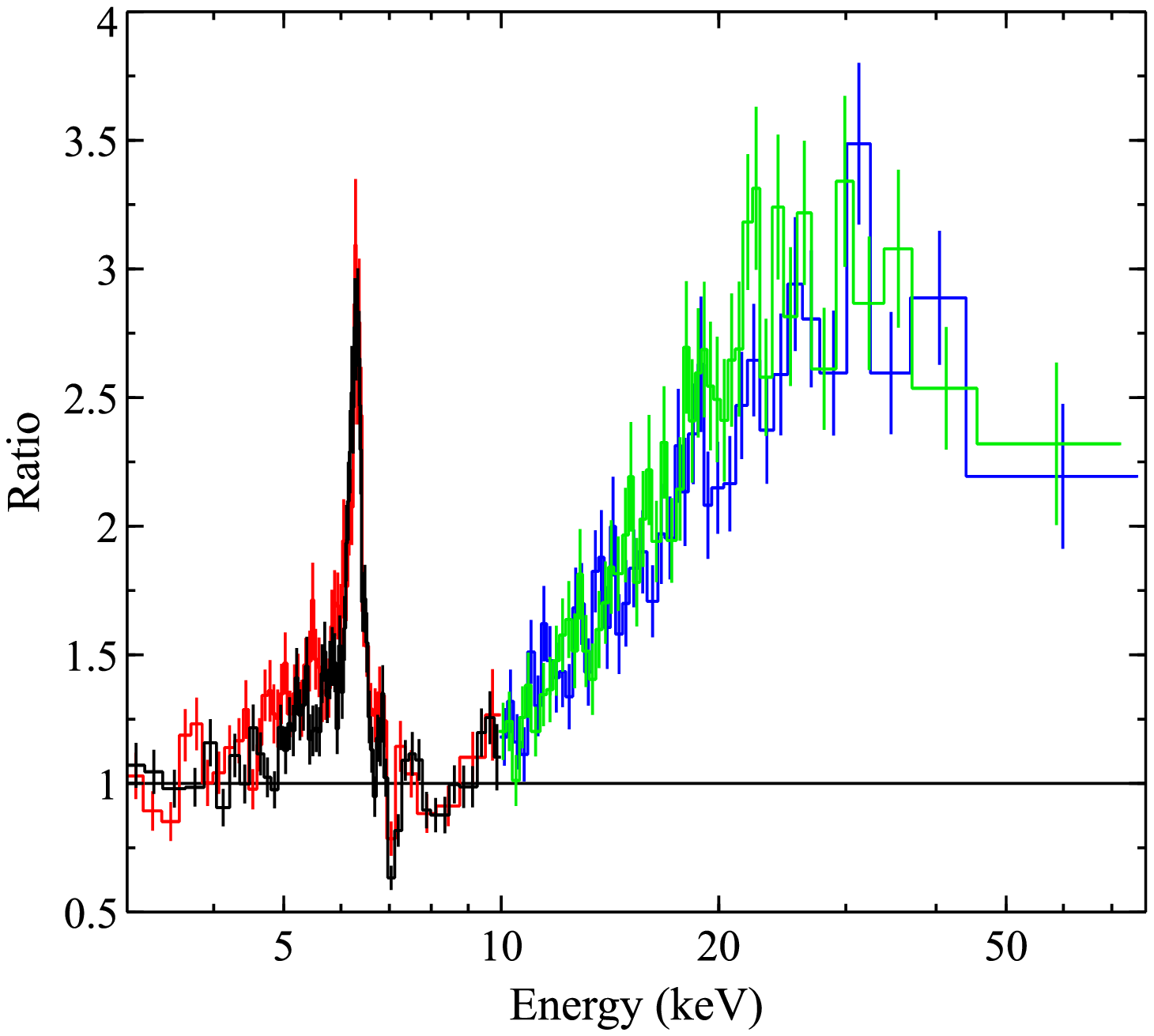}}
}
\end{center}
\caption{
\textit{Left panel:} time-averaged spectra from our coordinated
\nustar+\xmm\ observation of \iras, unfolded through a model that is constant with
energy. Data from the \epicpn\ and \epicmos\ detectors on \xmm\ and the FPMA
and FPMB modules on \nustar\ are shown in black, red, green and blue,
respectively, and for clarity we show the \xmm\ data below 10\,\kev, and the
\nustar\ data above 10\,\kev. At low energies, the emission is dominated by
photoionised plasmas rather than the intrinsic AGN emission, so we focus our
analysis above 3\,keV. \textit{Right panel:} residuals to a simple $\Gamma = 2.0$
powerlaw continuum, modified by partially covering neutral absorption, and
applied to the 3--4, 8--10 and 50--78\,\kev\ energy ranges. The hallmarks of
relativistic reflection from the inner accretion disk, \ie a broadened iron line at
$\sim$6\,\kev\ and a strong Compton hump at $\sim$30\,keV are seen. The data
in both panels have been rebinned for visual purposes.}
\label{fig_spec}
\end{figure*}

%\begin{figure*}
%\hspace*{-0.5cm}
%\epsscale{0.55}
%\plotone{./figs/iras13197_eeuf_XN1.eps}
%\hspace{0.5cm}
%\plotone{./figs/iras13197_ratio_po_XN1.eps}
%\caption{\textit{Left panel:} time-averaged spectra from our coordinated
%\nustar+\xmm\ observation of \iras, unfolded through a model that is constant with
%energy. Data from the \epicpn\ and \epicmos\ detectors on \xmm\ and the FPMA
%and FPMB modules on \nustar\ are shown in black, red, green and blue,
%respectively, and for clarity we show the \xmm\ data below 10\,\kev, and the
%\nustar\ data above 10\,\kev. At low energies, the emission is dominated by
%photoionised plasmas rather than the intrinsic AGN emission, so we focus our
%analysis above 3\,keV. \textit{Right panel:} residuals to a simple $\Gamma = 2.0$
%powerlaw continuum, modified by partially covering neutral absorption, and
%applied to the 3--4, 8--10 and 50--78\,\kev\ energy ranges. The hallmarks of
%relativistic reflection from the inner accretion disk, \ie a broadened iron line at
%$\sim$6\,\kev\ and a strong Compton hump at $\sim$30\,keV are seen. The data
%in both panels have been rebinned for visual purposes.}
%\vspace{0.3cm}
%\label{fig_spec}
%\end{figure*}

We show the full broadband spectrum observed in the left panel of Figure
\ref{fig_spec}. \iras\ is known to be quite heavily absorbed, and during this epoch
the hard AGN component only dominates above 3\,keV; at lower energies the
observed emission is dominated by a combination of photoionised gas and
thermal plasma emission (\citealt{Miniutti07iras}). Since we are interested in the
emission from the AGN in this work, we limit our analysis to energies above 3\,keV.
The right panel of Figure \ref{fig_spec} shows the data/model ratio of the combined
\xmm+\nustar\ dataset to a fairly typical $\Gamma = 2$ AGN continuum, modified
by a simple, partially covering neutral absorber (modelled with \tbnewpcf, a partially
covering version of the \tbnew\ absorption code), fit to the 3--4, 8--10 and
50--78\,keV bands where the primary AGN continuum would be expected to
dominate. The column density and covering factor are $\sim$7$ \times
10^{23}$\,\pcmsq\ and $\sim$0.94, respectively; adopting a different continuum
slope pivots the residuals (and changes the absorption parameters) but does not
change their overall shape. A strong, broad feature is clearly seen in the iron
bandpass, and a strong excess of emission is observed above 10\,keV, similar to
the broad iron K emission and strong hard excess previously reported for this source
(\citealt{Dadina04, Miniutti07iras, Tatum13}). In addition to these broad features, a
narrow core to the iron emission at 6.4\,keV and a narrow absorption feature just
above 7\,keV are both evident. \iras\ is also known to exhibit absorption from ionized
iron (\citealt{Miniutti07iras}), and there is evidence for a second narrow absorption
line at $\sim$6.7\,keV (this is not so visually obvious, but is also required by the data;
see below); given the observed energies and the known redshift of \iras\ ($z=0.0165$),
these absorption lines are likely associated with \fexxv\ and \fexxvi\ from an outflowing
disk wind. Prior studies have indicated an outflow velocity of $v_{\rm{out}} \sim
5000$\,\kmps\ (\citealt{Miniutti07iras}).

In order to fit the AGN spectrum, we therefore construct models consisting of a
partially covering neutral absorber, a powerlaw/Compton scattered continuum,
reflection from the accretion disk, reflection from more distant material, and an
ionized absorber. We assume the distant reflection arises from the same structure
that results in the neutral absorption, which is likely the putative torus invoked in
AGN unification models. We therefore group the model components into `distant'
components (neutral absorption, distant reflection) and `inner' components
(primary continuum, disk reflection, ionised absorption), and set the models up such
that the neutral absorption acts on all of the inner components, but does not act on
the distant reflection.

\begin{table*}
  \caption{Results obtained for the free parameters in the various lamppost
  reflection models fit to epoch XN1.}
\begin{center}
\begin{tabular}{c c c c c c c c c}
\hline
\hline
\\[-0.2cm]
Model Component & \multicolumn{2}{c}{Parameter} & \multicolumn{4}{c}{Model} \\
\\[-0.25cm]
& & & 1 & 2 & 3 & 4 \\
\\[-0.3cm]
\hline
\hline
\\[-0.1cm]
\tbnewpcf\ & \nh\ & [$10^{23}$ cm$^{-2}$] & $8.1 \pm 0.4$ & $7.6 \pm 0.4$ & $6.2^{+0.3}_{-0.5}$ & $6.4^{+0.2}_{-0.4}$ \\
\\[-0.3cm]
& $C_{\rm{f}}$ & [\%] & $>99.8$ & $98.7^{+1.1}_{-0.5}$ & $98.5 \pm 0.2$ & $99.2^{+0.1}_{-0.3}$ \\
\\[-0.3cm]
& $A_{\rm{Fe, distant}}$ & [solar] & =1 & =1 & =$A_{\rm{Fe, disc}}$ & =$A_{\rm{Fe, disc}}$ \\
\\
\relxilllp{\small{ (\_CP)}} & $\Gamma$ & & $2.16^{+0.10}_{-0.37}$ & $2.1^{+0.3}_{-0.2}$ & $2.2^{+0.1}_{-0.2}$ & $2.12^{+0.04}_{-0.32}$ \\
\\[-0.3cm]
& $E_{\rm{cut}}$ or $kT_{\rm{e}}$\tmark[a] & [keV] & $150^{+100}_{-40}$ & $100^{+300}_{-40}$ & $>110$ & $>48$ \\
\\[-0.3cm]
& $a^*$ & & $0.97^{+0.02}_{-0.04}$ & $0.92 \pm 0.03$ & $0.73^{+0.25}_{-0.32}$ & $0.72^{+0.23}_{-0.53}$ \\
\\[-0.3cm]
& $i$ & [\deg] & $44^{+2}_{-3}$ & $44^{+2}_{-3}$ & $61^{+2}_{-3}$ & $60^{+2}_{-3}$ \\
\\[-0.3cm]
& $h$ & [\rh] & $<1.4$ & $<1.4$ & $<1.8$ & $<1.8$ \\
\\[-0.3cm]
& \Rfrac\tmark[c] & & $5.1^{+3.3}_{-1.9}$ & $3.0^{+0.8}_{-0.4}$ & $1.7^{+1.7}_{-0.6}$ & $1.6^{+1.4}_{-0.7}$ \\
\\[-0.3cm]
& $\log\xi$ & $\log$[\ergcmps] & $<1.3$ & $<1.5$ & $<2.3$ & $<3.0$ \\
\\[-0.3cm]
& $A_{\rm{Fe, disc}}$\tmark[b] & [solar] & $>7.1$ & $>7.3$ & $2.3^{+0.2}_{-0.1}$ & $2.3^{+0.2}_{-0.1}$ \\
\\[-0.3cm]
& Norm & [$10^{-3}$] & $21.2^{+2.8}_{-9.5}$ & $10.3^{+17.0}_{-2.2}$ & $4.4^{+3.7}_{-1.2}$ & $3.3^{+4.1}_{-1.8}$ \\
\\
\xstar$_{\rm{abs}}$ & $\log\xi$ & $\log$[\ergcmps] & $3.46^{+0.13}_{-0.09}$ & $3.47^{+0.14}_{-0.10}$ & $3.51^{+0.04}_{-0.09}$ & $3.48 \pm 0.09$ \\
\\[-0.3cm]
& \nh\ & [$10^{23}$ cm$^{-2}$] & $0.6^{+0.4}_{-0.2}$ & $0.8^{+1.0}_{-0.3}$ & $2.4^{+1.1}_{-0.7}$ & $2.3^{+1.2}_{-0.6}$ \\
\\[-0.3cm]
& $v_{\rm{out}}$ & [\kmps] & $4900^{+600}_{-700}$ & $4800 \pm 700$ & $4700 \pm 700$ & $4700 \pm 700$ \\
\\
\xillver{\small{ (\_CP)}} & Norm & [$10^{-5}$] & $5.1^{+0.8}_{-0.3}$ & -- & -- & $3.3^{+0.4}_{-1.2}$ \\
\\[-0.3cm]
& $F_{\rm{dist/relx}}$\tmark[d] & & $0.19^{+0.03}_{-0.02}$ & -- & -- & $0.14 \pm 0.02$ \\
\\
\mytorus\ & $N_{\rm{H}}$ & [$10^{23}$\,\pcmsq] & -- & $22^{+16}_{-8}$ & $>32$ & -- \\
\\[-0.3cm]
& Norm & [$10^{-2}$] & -- & $1.5^{+0.7}_{-0.4}$ & $0.9^{+0.5}_{-0.2}$ & -- \\
\\[-0.3cm]
& $F_{\rm{dist/relx}}$\tmark[d] & & -- & $0.49^{+0.20}_{-0.37}$ & $0.19^{+0.03}_{-0.04}$ & -- \\
\\[-0.3cm]
\hline
\\[-0.2cm]
\chisq/DoF & & & 794/723 & 796/722 & 789/722 & 787/723 \\
\\[-0.3cm]
\hline
\hline
\end{tabular}
\label{tab_param_XN1}
\end{center}
\flushleft
$^a$ $E_{\rm{cut}}$ is limited to $\leq$1000\,keV following \cite{Garcia15}, and
$kT_{\rm{e}}$ is limited by the bounds of the {\scriptsize RELXILL\_CP} grid to
$20 \leq kT_{\rm{e}} \leq 400$\,keV. \\
$^b$ The iron abundance is limited by the bounds of the {\scriptsize XSTAR} grid
used here to $A_{\rm{Fe}} \leq 10.0$. \\
$^c$ \Rfrac\ is calculated self-consistently for a simple lamppost geometry
from $a^*$ and $h$. Errors represent the range of values permitted by varying these
parameters within their 90\% uncertainties. \\
$^d$ Ratio of the observed fluxes from the distant reflector and the
\relxilllp{\small{ (\_CP)}} components in the 20--40\,keV band.
\vspace{0.4cm}
\end{table*}

\subsubsection{Model 1: Basic Approach}

In this work, we model the innermost regions of the AGN with a simple lamppost
geometry using the \relxill\ family of disk reflection models (\citealt{relxill}), which
merges the \xillver\ family of reflection models with the \relconv\ model for the
relativistic effects relevant for regions close to a black hole (\citealt{relconv}). The
lamppost geometry treats the X-ray emitting region as a point source situated on
the rotation axis of the black hole. While this is obviously an idealised geometric 
approximation, it provides a simple parametrization for the reflected emission and
allows us to exclude low-spin, reflection-dominated solutions that are unphysical
for thin-disk accretion (\citealt{Dauser14}). We start by using the \relxilllp\ model,
which treats the illuminating continuum as a simple powerlaw with a high-energy
exponential cutoff. This model incorporates both the primary continuum and the
reflected emission from the accretion disk, and self-consistently determines both
the radial emissivity profile for the disk and the relative strength of the reflected
emission (\Rfrac, defined to be the ratio between the continuum fluxes seen by
the disk and by the observer in the latest versions of the \relxilllp\ model; see
\citealt{relxill_norm}) based on the spin of the black hole ($a^{*}$; we assume the
disk extends into the innermost stable circular orbit) and the height of the X-ray
source ($h$; see also \citealt{Wilkins12}). In order to ensure that the source is
always required to be outside the event horizon, we fit the model with $h$ in units
of the vertical event horizon radius (\rh, which varies from $1-2$\,\rg\ for
maximally-rotating to non-rotating black holes, where \rg\ = $GM/c^{2}$ is the
gravitational radius). The other key free parameters are the photon index and
high-energy cutoff of the illuminating continuum ($\Gamma$, $E_{\rm{cut}}$), the
iron abundance ($A_{\rm{Fe}}$), inclination ($i$) and ionisation parameter ($\xi$)
of the accretion disk. Here, the ionisation parameter is defined as $\xi =
4{\pi}F_{\rm{X}}/n$, where $F_{\rm{X}}$ is the incident X-ray flux, and $n$ is the
density of the material.

The neutral absorption is again modelled with \tbnewpcf, and we initially treat the
distant reflection with an unblurred \xillver\ component. \tbnewpcf\ implicitly
assumes that the absorber has solar abundances, and only has the column density
($N_{\rm{H}}$) and the covering fraction ($C_{\rm{f}}$) as free parameters (we
assume the neutral absorber is at the redshift of the host galaxy). The \xillver\ model
assumes a simple slab geometry for the reflector, and its key free parameters are
again the photon index and high-energy cutoff of the illuminating continuum, the
ionization parameter and the iron abundance of the reflecting medium, and the
inclination of the slab. Given the assumed association between the distant reflector
and the neutral absorber, we fix the iron abundance for the \xillver\ component to
solar. The parameters for the illuminating continuum are assumed to be the same as
those for the \relxilllp\ component, and we fix the inclination to 45\deg, since the fits
are largely insensitive to this parameter. In order to provide the reader with an
indication of the relative contribution of the distant reflection, in addition to their 
normalisations we also present the ratio of the observed fluxes for the
distant reflector and the \relxilllp\ component ($F_{\rm{dist/relx}}$) in the 20--40\,keV
band, roughly where the Compton hump peaks; the effects of photoelectric absorption
from the neutral absorber are also reduced in this band. In addition to treating the
photoelectric absorption from the neutral absorber, as the column along our line of
sight to \iras\ can be quite large (see Section \ref{sec_oldobs}) we also account for
flux losses due to Compton-scattering in the intervening material by including a \cabs\
component with the column set to be equal to that of the neutral absorber. 

Finally, we model the ionised absorption with the \xstar\ photoionisation code,
computing a custom grid of absorption models. We allow the column density, the
ionisation parameter, the outflow velocity and the iron abundance of the absorber
to be varied as free parameters. The other abundances are assumed to be solar,
and the absorbing medium is assumed to be illuminated by a $\Gamma = 2$
powerlaw for the model calculation. Initial modelling of the two absorption lines with
Gaussian features (both lines are significant; the fit without any absorption
features included is improved by $\Delta\chi^{2} = 111$ for three additional free
parameters when the \fexxvi\ line is added, and by a further $\Delta\chi^{2} = 27$
for one more free parameter when the \fexxv\ line is added, assuming the same
outflow velocity and broadening for the two lines) suggested line broadening of
$\sigma \sim 0.06$\,keV, so we set the turbulent velocity to 1750\,\kmps. Since we
assume the outflow comes from the accretion disk, we link the iron abundance of
the ionised absorption to that of the disk reflection (\relxilllp) component in our
analysis. The final form of this initial model, which we refer to as Model 1, is:
\tbnew\ $\times ~ ($ \xillver\ $+ ~ ($ \tbnewpcf\ $\times$ \xstar\ $\times$ \cabs\
$\times$ \relxilllp\ $))$.

\begin{figure}
\begin{center}
%\hspace*{-0.5cm}
\rotatebox{0}{
{\includegraphics[width=235pt]{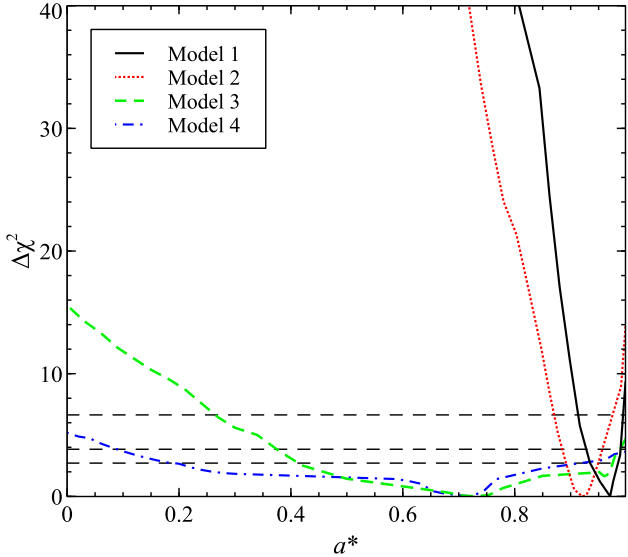}}
}
\end{center}
\caption{
The $\Delta$\chisq\ confidence contours for the black hole spin obtained
for Models 1--4 applied to epoch XN1. The horizontal dashed lines represent the
90, 95 and 99\% confidence levels for a single parameter of interest.}
\label{fig_spin_XN1}
\end{figure}

Applying this model to the combined 3--79\,keV spectrum from XN1 results in a
good fit, with \chisq\ = 794 for 723 degrees of freedom (DoF). The parameter values
obtained are given in Table \ref{tab_param_XN1}. The neutral absorption column
local to \iras\ is found to be rather large ($N_{\rm{H}} = [8.1 \pm 0.4] \times
10^{23}$\,\pcmsq), but the fits still imply a very strong disk reflection contribution
(\Rfrac $\sim$ 5; for reference, a thin accretion disk in Newtonian gravity
should give \Rfrac $\sim$ 1), broadly similar to that inferred previously
(\citealt{Miniutti07iras}). For a standard thin accretion disk, strong gravitational
lightbending is required to produce such strong reflection (\eg\ \citealt{lightbending}).
This in turn requires a very compact X-ray source in close vicinity to a rapidly rotating
black hole, such that the majority of the emission from the X-ray source is bent onto
the disk, rather than escaping to the observer directly (see also \citealt{Dauser14,
Parker14mrk}). The confidence contour for the black hole spin is shown in Figure
\ref{fig_spin_XN1}; we find $a^{*} = 0.97^{+0.01}_{-0.04}$ for this model. We also
find that, as expected, the height of the X-ray source is constrained to be very small:
$h \leq 1.6$\,\rg\ (converting from \rh\ based on the spin constraints).

%\begin{figure}
%\hspace*{-0.5cm}
%\epsscale{1.05}
%\plotone{./figs/iras13197_XN1_spin_contours_cabs.eps}
%\caption{The $\Delta$\chisq\ confidence contours for the black hole spin obtained
%for Models 1--4 applied to epoch XN1. The horizontal dashed lines represent the
%90, 95 and 99\% confidence levels for a single parameter of interest.}
%\vspace{0.3cm}
%\label{fig_spin_XN1}
%\end{figure}

\begin{figure*}
\begin{center}
\hspace*{-0.3cm}
\rotatebox{0}{
{\includegraphics[width=235pt]{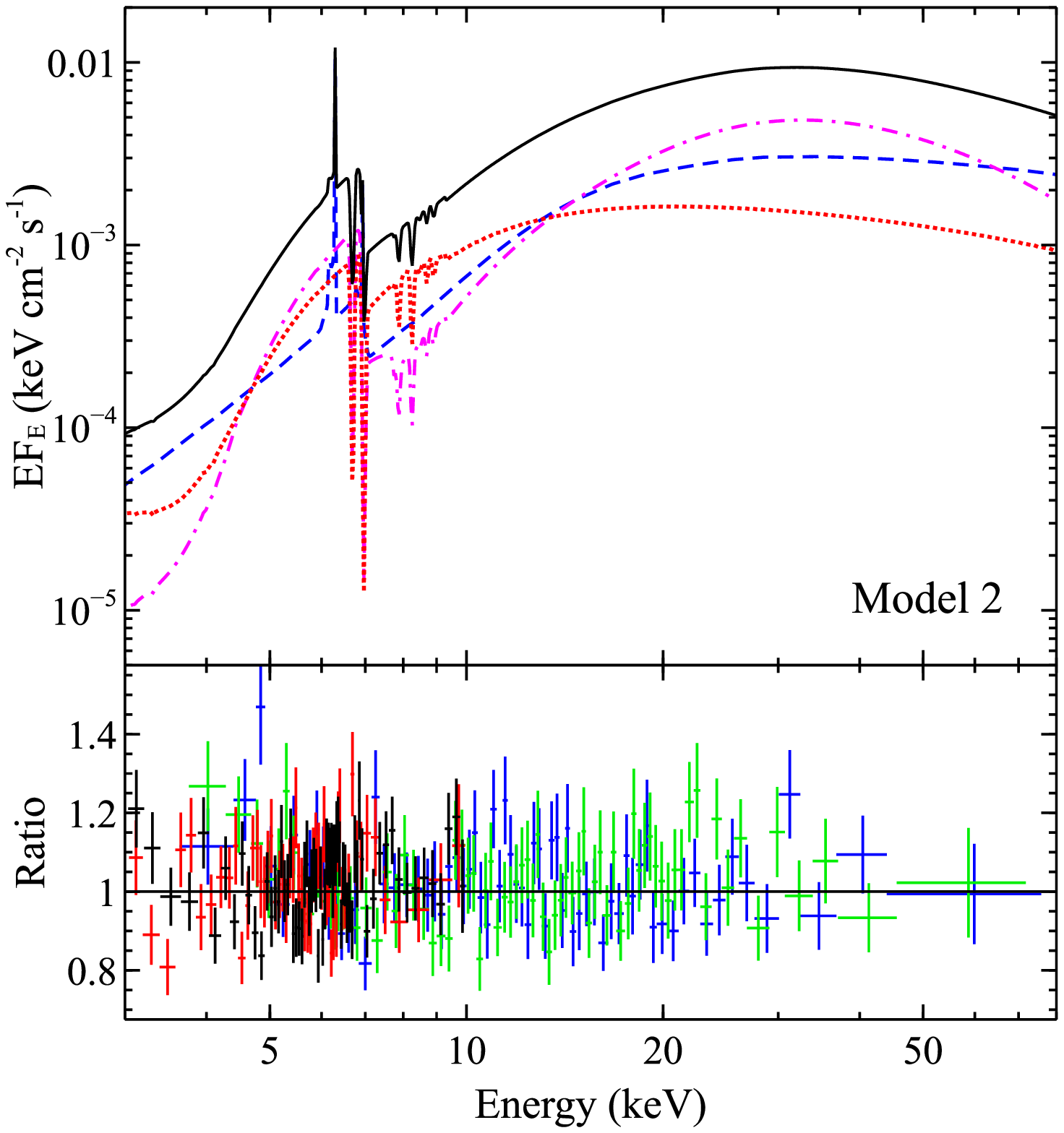}}
}
\hspace*{0.5cm}
\rotatebox{0}{
{\includegraphics[width=235pt]{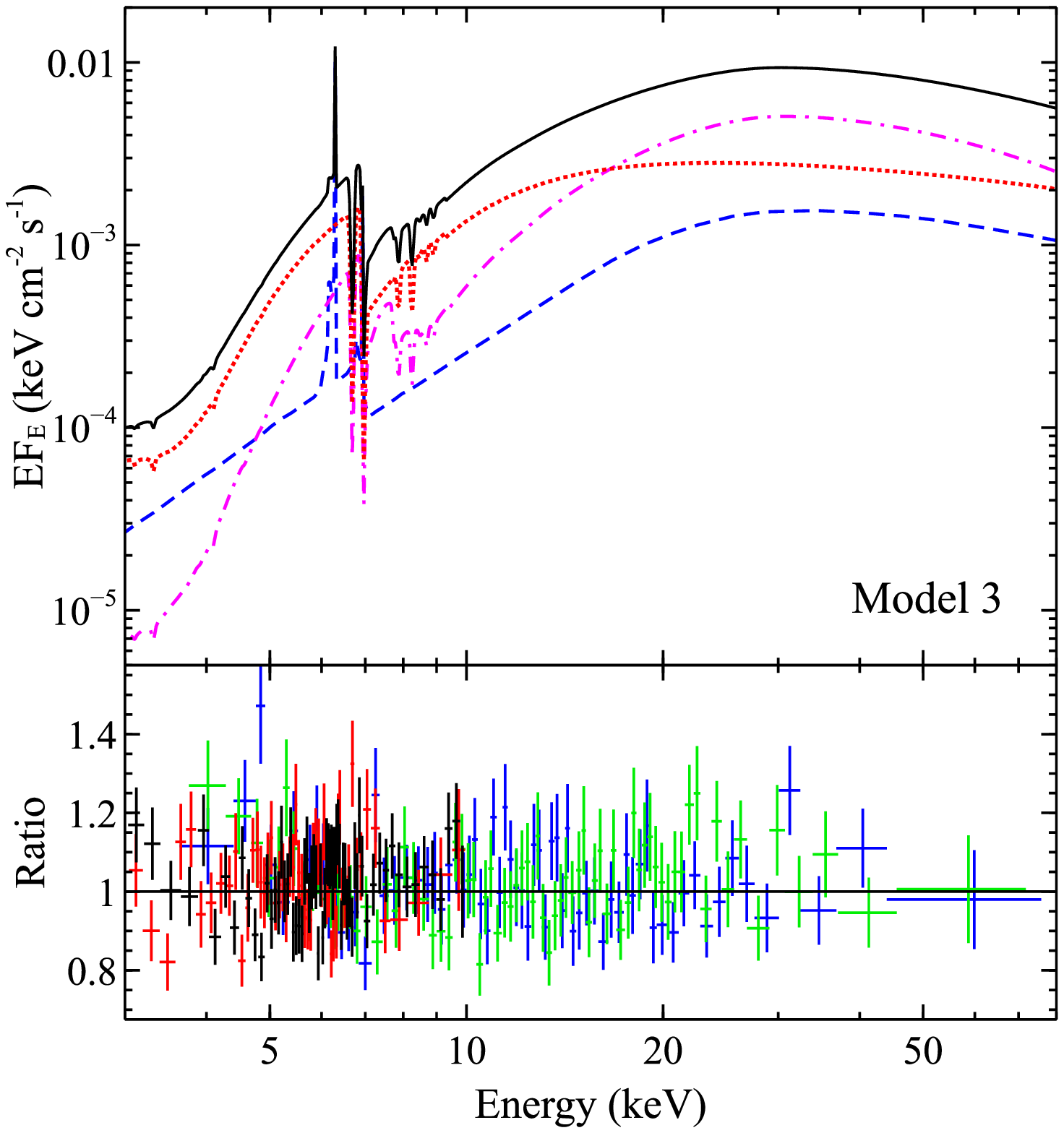}}
}
\end{center}
\caption{
A comparison of the best-fit models for epoch XN1 assuming a torus
geometry for the distant reprocessor: \textit{left panels:} a solar iron abundance
assumed for the distant model components; \textit{right panels:} assuming a
common iron abundance for all AGN model components. \textit{Top panels:} the
relative contributions of the different model components, with the total model
shown in solid black, the powerlaw continuum in dotted red, the inner disk reflection
in dash-dot magenta and distant reflection in dashed blue. \textit{Bottom panels:}
data/model ratios for these models. The colour scheme for these panels is the same
as Figure \ref{fig_spec}, and again the data have been rebinned for visual purposes.}
\label{fig_modcomp}
\end{figure*}

%\begin{figure*}
%\hspace*{-0.5cm}
%\epsscale{0.54}
%\plotone{./figs/iras13197_XN1_eemrat_RXMT_distsolar_cabs.eps}
%\hspace{0.6cm}
%\plotone{./figs/iras13197_XN1_eemrat_RXMT_absFelink_cabs.eps}
%\caption{A comparison of the best-fit models for epoch XN1 assuming a torus
%geometry for the distant reprocessor: \textit{left panels:} a solar iron abundance
%assumed for the distant model components; \textit{right panels:} assuming a
%common iron abundance for all AGN model components. \textit{Top panels:} the
%relative contributions of the different model components, with the total model is
%shown in black, the powerlaw continuum in red, the inner disk reflection in
%magenta and distant reflection in blue. \textit{Bottom panels:} data/model ratios
%for these models. The colour scheme for these panels is the same as Figure
%\ref{fig_spec}, and again the data have been rebinned for visual purposes.}
%\vspace{0.3cm}
%\label{fig_modcomp}
%\end{figure*}

\subsubsection{Model 2: MYTorus}

The \xillver\ model used to account for the distant reflection contribution in Model
1 assumes a simple slab geometry for the reprocessor, which is also treated as being
highly optically-thick. While convenient, this is unlikely to be physically realistic for the
distant reprocessor in an AGN, which is expected to have a torus-like geometry and a
finite column of material. Different geometries can lead to differences in the reflected
spectra (\eg\ \citealt{Brightman15}). In order to determine whether this could have any
influence on the results obtained for the disk reflection, we construct a second model
(hereafter Model 2) in which the \xillver\ component is replaced by \mytorus\
(\citealt{mytorus}). This self-consistently computes the absorption and reprocessed
emission from a neutral torus surrounding a central illuminating X-ray source.
\mytorus\ assumes a doughnut-like geometry such that the line-of-sight column
density varies with viewing angle, up to a maximum equatorial column density. The
basic version of \mytorus\ also assumes the illuminating continuum to be a powerlaw
(up to a termination energy of 500\,keV, which is outside the \nustar\ bandpass), that
the torus has a solar composition (based on the solar abundances reported in
\citealt{Anders89}), and that the torus has an opening of 30\deg\ from the equatorial
plane. The key \mytorus\ parameters are the equatorial column density, the angle at
which the torus is viewed, and the continuum photon index.

The 30\deg\ opening angle assumed for the torus is equivalent to a covering factor
of $\Omega = 2\pi$, or in fractional terms $\Omega_{\rm{f}} = 0.5$ (where
$\Omega_{\rm{f}} = \Omega/4\pi$). Although this covering factor is fixed in the
available \mytorus\ models, the assumed value is quite reasonable for \iras.
Correcting for losses from both photoelectric absorption and Compton scattering due
to the neutral line-of-sight absorber, we estimate the intrinsic 2--10\,keV luminosity
to be  $\sim$1.5 $\times$ $10^{43}$\,\ergps\ during epoch XN1. Based on the
correlation between the covering factor of the distant reprocessor and the intrinsic
2--10\,keV X-ray luminosity presented by \cite{Brightman15} we find an expected
covering factor of $\Omega_{\rm{f}} \sim0.5$.

The \mytorus\ model is separated into three different components: the absorption
from the torus, the continuum component to the reprocessed emission, and the
reprocessed line emission. The line component only includes Fe K$\alpha$ and
K$\beta$, but these are the key transitions relevant to this work. In order to
continue allowing for the possibility of partial covering, we still model the neutral
absorption with \tbnewpcf, but we replace the \xillver\ component with a
combination of the two reprocessed \mytorus\ components (lines plus continuum).
All parameters are required to be the same for both, but we additionally multiply
the line emission with an energy-independent constant factor of 0.68 (the
ratio of the \citealt{Anders89} and \citealt{Grevesse98} solar iron abundances) to
approximately account for the differences in the abundance sets used for
\mytorus\ and \xillver/\xstar, since the iron abundance isn't currently a free
parameter in \mytorus.\footnote{Multiplying the line component by a constant is
only a rough approximation for a different iron abundance as the strength of the
iron absorption edge should also vary correspondingly; see the {\scriptsize
MYTORUS} manual for caveats.} Since we are only using the reprocessed
\mytorus\ components, the fits are not strongly sensitive to the viewing angle for
the torus. However, as the neutral column varies from epoch to epoch (see below),
we assume that we are viewing the system fairly close to the edge of the torus,
and set this to 65\deg. Finally, we link the photon index to that of the \relxilllp\
component.

This model provides a similarly good fit to the data as Model 1 (see Table
\ref{tab_param_XN1}). Critically, although there are some minor quantitative
changes to some of the parameters, the results are generally similar to Model 1.
The reflection fraction is still high (\Rfrac $\sim$ 3), requiring a strong degree of
lightbending, and again a compact X-ray source and a rapidly rotating black hole.
In this case, we find $a^{*} = 0.92 \pm 0.03$ (see Figure \ref{fig_spin_XN1}). The
key results for the disk reflection do not depend strongly on the precise modelling
of the distant reflection. The best-fit \mytorus\ column density suggests that while
the neutral absorption along our line-of-sight is Compton-thin, the absorbing
medium does become Compton-thick closer to the equatorial plane.

\subsubsection{Model 3: Common Iron Abundance}

One issue of note with both Models 1 and 2 is that the iron abundance of the disk
is found to be very high. Similar iron abundances have been found in other AGN,
\eg\ 1H\,0707-495 (\citealt{FabZog09}) and IRAS\,13224--3809
(\citealt{Fabian13iras}). However, in this case the iron abundance of the disk
strongly contrasts the solar abundance assumed for the distant model
components (neutral absorption, distant reflection) which play a significant role in
shaping the observed spectrum (in contrast to \iras, both 1H\,0707--495 and
IRAS\,13224--3809 are unobscured systems). This is similar to results recently
reported by \cite{Xu17iras} for the Seyfert 2 galaxy IRAS\,05189--2524.

In order to investigate this issue further, we construct a third model in which all the
distant and the disk components have a common iron abundance (hereafter Model
3), mimicking the approach taken in \cite{Miniutti07iras}. We continue to model the
distant reflection with \mytorus\ here, but stress that again similar results are seen if
we replace this with \xillver. Since the \tbnewpcf\ model does not include iron
abundance as a free parameter, in order to account for the neutral absorption we
replace this with a version of \tbnew\ in which the iron abundance can be varied,
convolved with a \partcov\ component to continue allowing for the possibility of
partial covering (for simplicity, we still refer to this combination as \tbnewpcf\ in Table
\ref{tab_param_XN1}). We then link all the iron abundance parameters together,
along with the constant multiplicative factor applied to the \mytorus\ line emission
(continuing to include the additional factor of 0.68 to account for the different solar
iron abundance assumed by \mytorus).

This model provides a similarly good fit to the data as both Models 1 and 2
(again, see Table \ref{tab_param_XN1}). However, there are some significant
differences of note between the results obtained. The iron abundance is still
super-solar ($A_{\rm{Fe}} = 2.3^{+0.2}_{-0.1}$), but not extremely so, and is
(unsurprisingly) intermediate to the solar abundance assumed for the distant
components and the high abundance found for the disk components in the
previous models. This allows the neutral absorption to account for more of the
strong drop in the spectrum above $\sim$7\,keV (the iron edge) and produce
more curvature in the observed continuum above 10\,keV, which reduces the
requirement for the reflection model to account for these features with a large
reflection fraction. In turn, this means the requirement for a rapidly rotating black
hole is no longer as strong, and so the parameter constraints are
correspondingly looser. In this case, we find that $a^* = 0.73^{+0.25}_{-0.32}$
(see Figure \ref{fig_spin_XN1}). However, the best-fit reflection fraction is still
greater than unity, so the X-ray source is still required to be relatively compact.
The increased iron abundance in the absorber also means that the blue wing of
the iron line in the disk reflection component is no longer required to be at a low
enough energy to help produce the $\sim$7\,keV spectral drop, which results in
a higher best-fit inclination. Lastly, the stronger line emission relative to the
continuum in the distant reflection means this component cannot account for as
much of the observed emission at $\sim$3--4\,keV as in the previous models,
and so the column of the neutral absorber is now significantly lower to
compensate. A comparison of Models 2 and 3 is shown in Figure
\ref{fig_modcomp}.

\subsubsection{Model 4: Compton-Scattered Continuum}
\label{sec_compt}

All three of the models presented so far have treated the primary continuum as a
simple powerlaw with an exponential high-energy cutoff. The reflection components
used so far have also been calculated assuming the illuminating continuum to be a
simple powerlaw continuum either with (\relxilllp, \xillver) or without (\mytorus) an
exponential high-energy cutoff. However, the nature of the primary X-ray continuum
is widely expected to be Compton up-scattering of low-energy photons from the
accretion disk by hot electrons for most AGN (\eg\ \citealt{Haardt91}). While a cutoff
powerlaw is often a reasonable approximation for a real Compton-scattered
continuum, there are subtle differences between them. The former is constantly
curving across all energies, while the latter is more powerlaw-like until it rolls over
with a sharper cutoff (\eg\ \citealt{Zdziarski03, Fabian15, Fuerst16}).

Recently, versions of the \xillver\ family of models have been calculated using
a realistic Compton-scattered continuum for the illuminating continuum (hereafter
\xillvercp, \etc). Therefore, we construct one more model utilizing these new
versions to investigate what effect the differences between the assumed continuum
forms might have on the results for the disk reflection (Model 4), replacing \relxilllp\
with \relxilllpcp, and since the results with \mytorus\ and \xillver\ have been
consistent, replacing \mytorus\ with \xillvercp\ for convenience.\footnote{Table
models for \mytorus\ have also recently been computed for a realistic
Compton-scattered illuminating continuum, but these use the {\scriptsize COMPTT}
thermal Comptonization model (\citealt{comptt}) which has a different
parametrisation to \nthcomp, and these grids do not directly include the electron
temperature as a free parameter.} \xillvercp\ and \relxilllpcp\ have been calculated
using the \nthcomp\ thermal Comptonization model (\citealt{nthcomp1, nthcomp2})
as the input continuum, which is primarily parametrized by the photon index of the
continuum below the cutoff, and the electron temperature ($kT_{\rm{e}}$). The key
parameters for \xillvercp\ and \relxilllpcp\ are therefore essentially the same as for
\xillver\ and \relxilllp, except that the high-energy cutoff parameter has been
replaced with $kT_{\rm{e}}$. As before, we assume the same illuminating
continuum parameters for both the \relxilllpcp\ and \xillvercp\ components, and we
also continue with the approach taken in Model 3 and assume a common iron
abundance for all AGN components.

Model 4 also provides a similarly good fit to all the previous models considered (as
before, see Table \ref{tab_param_XN1}). The different curvature in the primary
continuum allows in this case for lower values of \Rfrac, as the sharper
curvature in the primary continuum allows this component to account for a bit more
of the observed high-energy curvature by decreasing the electron temperature, a
degeneracy that is naturally exacerbated by the strong absorption in the system.
The formal constraint on the spin therefore weakens a little further in comparison to
Model 3; here we find $a^* = 0.72^{+0.23}_{-0.53}$. The rest of the parameters
remain broadly similar to the results with Model 3.

\begin{figure}
\begin{center}
%\hspace*{-0.5cm}
\rotatebox{0}{
{\includegraphics[width=235pt]{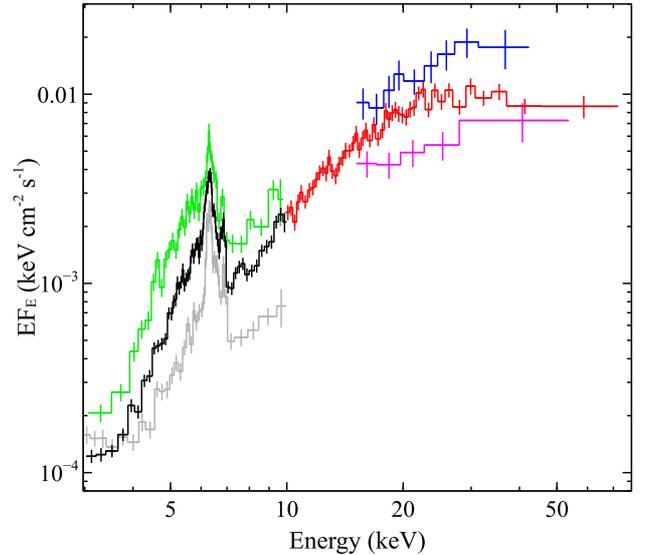}}
}
\end{center}
\caption{
Multi-epoch spectra observed from \iras. As in Figure \ref{fig_spec}, the data have
been unfolded through a model that is constant with energy, and we only show
epochs S1 (high flux datasets), S2 (low flux datasets) and XN1 (medium flux datasets),
as epochs X1 and X2 are similar to epoch S1. For clarity, we only show the \epicpn\
data from \xmm\ (black; below 10\,keV), the FPMA data from \nustar\ (red; above
10\,keV), and the data from the front-illuminated XIS detectors (S1: green, S2: grey)
and the PIN detector (S1: blue, S2: magenta) from \suzaku. Epochs X1, X2 and S1
show lower levels of line-of-sight absorption than XN1, while epoch S2 has stronger
absorption.}
\label{fig_spec_all}
\end{figure}

%\begin{figure}
%\hspace*{-0.5cm}
%\epsscale{1.05}
%\plotone{./figs/iras13197_ME_eeuf.eps}
%\caption{Multi-epoch spectra observed from \iras. As in Figure \ref{fig_spec}, the
%data have been unfolded through a model that is constant with energy, and we
%only show epochs S1, S2 and XN1, as epochs X1 and X2 are similar to epoch
%S1. For clarity, we only show the \epicpn\ data from \xmm\ (black), the FPMA data
%from \nustar\ (red), and the data from the front-illuminated XIS detectors (S1:
%green, S2: grey) and the PIN detector (S1: blue, S2: magenta) from \suzaku.
%Epochs X1, X2 and S1 show lower levels of line-of-sight absorption than XN1,
%while epoch S2 has stronger absorption.}
%\vspace{0.3cm}
%\label{fig_spec_all}
%\end{figure}

Finally, although we do not present the full results to these fits as additional models
here, we note that if we repeat the fit assuming the distant model components to
have a solar abundance, but assuming a Comptonized continuum instead of a cutoff
powerlaw, we find practically identical results to those presented for Model 1. In
this scenario, the precise form of the continuum does not have any significant
effect. We also note that allowing the iron abundances for both the distant and the
disk components to vary independently does not result in any significant statistical
improvement over the common iron abundance scenario.

\subsection{Multi-Epoch Analysis}
\label{sec_oldobs}

In order to obtain the most robust constraints on the key inner disk parameters, in
addition to our analysis of epoch XN1 we also undertake a multi-epoch analysis
of all the observations included in Table \ref{tab_obs}. As with XN1, we only analyse
the data above 3\,keV. A comparison of some of the spectra from these observations
is shown in Figure \ref{fig_spec_all}. During the first three of these archival
observations (X1, X2, S1), \iras\ was less absorbed than during epoch XN1, and the
spectra of these three datasets over their common 3--10\,keV bandpass were
extremely similar. In contrast, during the last of the archival observations (S2), \iras\
was significantly more absorbed than XN1. Despite the changing absorption, the
\suzaku\ PIN detections from S1 and S2 suggest that the intrinsic variations are
relatively minor; the observed 20--40\,keV fluxes only vary by a factor of $\sim$2.

\begin{table*}
  \caption{Results obtained for the free parameters in our multi-epoch lamppost
  reflection model fit.}
\begin{center}
\begin{tabular}{c c c c c c c c c}
\hline
\hline
\\[-0.2cm]
Model Component & \multicolumn{2}{c}{Parameter} & Global & \multicolumn{5}{c}{Epoch} \\
\\[-0.25cm]
& & & & X1 & X2 & S1 & S2 & XN1 \\
\\[-0.3cm]
\hline
\hline
\\[-0.2cm]
\multicolumn{9}{c}{Intrinsic variations: variable RELXILLLP normalisation, constant $h$} \\
\\[-0.2cm]
\tbnewpcf\ & \nh\ & [$10^{23}$\,\pcmsq] & & $5.3 \pm 0.3$ & =X1 & =X1 & $10.0^{+0.8}_{-0.9}$ & $6.3^{+0.5}_{-0.3}$ \\
\\[-0.3cm]
& $C_{\rm{f}}$ & [\%] & & $98.7 \pm 0.2$ & =X1 & =X1 & $96.7^{+0.9}_{-0.8}$ & $98.6^{+0.3}_{-0.2}$ \\
\\
\relxilllpcp\ & $\Gamma$ & & $1.75^{+0.13}_{-0.05}$ \\
\\[-0.3cm]
& $kT_{\rm{e}}$\tmark[a] & [keV] & $>33$ \\
\\[-0.3cm]
& $a^*$ & & $>0.92$ \\
\\[-0.3cm]
& $i$ & [\deg] & $59^{+5}_{-8}$ \\
\\[-0.3cm]
& $h$ & \rh\ & $<2.7$ \\
\\[-0.3cm]
& \Rfrac\tmark[b] & & $5.4^{+3.5}_{-2.4}$ \\
\\[-0.3cm]
& $\log\xi$ & $\log$[\ergcmps] & $3.40^{+0.10}_{-0.13}$ \\
\\[-0.3cm]
& $A_{\rm{Fe}}$ & [solar] & $2.2 \pm 0.2$ \\
\\[-0.3cm]
& Norm & $10^{-3}$ & & $2.3^{+2.0}_{-1.4}$ & $2.8^{+2.8}_{-1.7}$ & $3.1^{+3.3}_{-1.6}$ & $1.7^{+1.9}_{-0.9}$ & $2.1^{+2.0}_{-1.1}$ \\
\\
\xstar$_{\rm{abs}}$ & $\log\xi$ & $\log$[\ergcmps] & $3.46 \pm 0.06$ \\
\\[-0.3cm]
& \nh\ & [$10^{23}$ cm$^{-2}$] & & $1.7^{+1.0}_{-0.5}$ & $4.0^{+1.4}_{-1.0}$ & $6.0^{+3.0}_{-2.2}$ & $0.9^{+1.0}_{-0.5}$ & $1.9^{+0.6}_{-0.5}$ \\
\\[-0.3cm]
& $v_{\rm{out}}$ & \kmps & & $5900^{+1300}_{-1500}$ & $7200 \pm 600$ & $5400 \pm 900$ & $7800^{+3300}_{-2100}$ & $4700^{+800}_{-700}$ \\
\\
\xillvercp\ & Norm & [$10^{-5}$] & $2.4^{+0.7}_{-0.3}$ \\
\\[-0.3cm]
& $F_{\rm{dist/relx}}$\tmark[c] & & & $0.22^{+0.02}_{-0.04}$ & $0.18^{+0.02}_{-0.03}$ & $0.16^{+0.02}_{-0.03}$ & $0.44^{+0.07}_{-0.09}$ & $0.26^{+0.02}_{-0.04}$ \\
\\[-0.3cm]
\hline
\\[-0.2cm]
\chisq/DoF & & & 1673/1545 \\
\\[-0.3cm]
\hline
\hline
\\[-0.2cm]
\multicolumn{9}{c}{Geometric variations: constant RELXILLLP normalisation, variable $h$} \\
\\[-0.2cm]
\tbnewpcf\ & \nh\ & [$10^{23}$\,\pcmsq] & & $5.0 \pm 0.2$ & =X1 & =X1 & $9.2^{+0.8}_{-0.7}$ & $5.9 \pm 0.3$ \\
\\[-0.3cm]
& $C_{\rm{f}}$ & [\%] & & $98.7 \pm 0.2$ & =X1 & =X1 & $96.3^{+0.7}_{-0.8}$ & $98.4 \pm 0.2$ \\
\\
\relxilllpcp\ & $\Gamma$ & & $1.79^{+0.10}_{-0.04}$ \\
\\[-0.3cm]
& $kT_{\rm{e}}$\tmark[a] & [keV] & $<42$ \\
\\[-0.3cm]
& $a^*$ & & $>0.70$ \\
\\[-0.3cm]
& $i$ & [\deg] & $60^{+3}_{-4}$ \\
\\[-0.3cm]
& $h$ & \rh\ & & $4.6^{+2.6}_{-1.7}$ & $6.7^{+5.7}_{-3.1}$ & $9.7^{+13.0}_{-5.3}$ & $2.9^{+1.6}_{-0.7}$ & $3.9^{+1.7}_{-1.3}$ \\
\\[-0.3cm]
& \Rfrac\tmark[b] & & & $1.7 \pm 0.2$ & $1.5^{+0.1}_{-0.2}$ & $1.3 \pm 0.2$ & $2.2^{+0.4}_{-0.6}$ & $1.9^{+0.2}_{-0.3}$ \\
\\[-0.3cm]
& $\log\xi$ & $\log$[\ergcmps] & $3.0^{+0.2}_{-0.3}$ \\
\\[-0.3cm]
& $A_{\rm{Fe}}$ & [solar] & $2.6^{+0.1}_{-0.2}$ \\
\\[-0.3cm]
& Norm & $10^{-3}$ & $0.17^{+0.10}_{-0.03}$ \\
\\
\xstar$_{\rm{abs}}$ & $\log\xi$ & $\log$[\ergcmps] & $3.44 \pm 0.06$ \\
\\[-0.3cm]
& \nh\ & [$10^{23}$ cm$^{-2}$] & & $1.7^{+1.1}_{-0.6}$ & $3.9^{+1.2}_{-1.1}$ & $5.6^{+2.6}_{-2.0}$ & $0.9^{+0.9}_{-0.4}$ & $1.8 \pm 0.5$ \\
\\[-0.3cm]
& $v_{\rm{out}}$ & \kmps & & $5800^{+1200}_{-1500}$ & $7000 \pm 600$ & $5300^{+900}_{-800}$ & $7600^{+2500}_{-1700}$ & $4600 \pm 700$ \\
\\
\xillvercp\ & Norm & [$10^{-5}$] & $1.8^{+0.4}_{-0.1}$ \\
\\[-0.3cm]
& $F_{\rm{dist/relx}}$\tmark[c] & & & $0.14 \pm 0.02$ & $0.12 \pm 0.02$ & $0.11^{+0.02}_{-0.01}$ & $0.27^{+0.05}_{-0.04}$ & $0.16 \pm 0.02$ \\
\\[-0.3cm]
\hline
\\[-0.2cm]
\chisq/DoF & & & 1667/1545 \\
\\[-0.3cm]
\hline
\\[-0.2cm]
$F_{3-10}$\tmark[d] & \multicolumn{2}{c}{[$10^{-12}$\,\ergpcmsqps]} & & $2.64^{+0.06}_{-0.03}$ & $2.97 \pm 0.05$ & $3.15 \pm 0.10$ & $0.95^{+0.02}_{-0.03}$ & $1.88 \pm 0.03$ \\
\\[-0.3cm]
$F_{20-40}$\tmark[d] & \multicolumn{2}{c}{[$10^{-12}$\,\ergpcmsqps]} & & -- & -- & $14.5 \pm 0.8$ & $6.3 \pm 0.5$ & $9.7^{+0.3}_{-0.4}$ \\
\\[-0.3cm]
\hline
\hline
\end{tabular}
\label{tab_param_all}
\end{center}
%\vspace{0.1cm}
\flushleft
$^a$ $kT_{\rm{e}}$ is limited by the bounds of the {\scriptsize RELXILL\_CP} grid to
$20 \leq kT_{\rm{e}} \leq 400$\,keV. \\
$^b$ \Rfrac\ is calculated self-consistently for a simple lamppost geometry
from $a^*$ and $h$. Errors represent the range of values permitted by varying these
parameters within their 90\% uncertainties. \\
$^c$ Ratio of the observed fluxes from the \xillvercp\ and the \relxilllpcp\ components in
the 20--40\,keV band. \\
$^d$ Observed fluxes, not corrected for the line-of-sight absorption. These are
consistent for both models.
\vspace{0.3cm}
\end{table*}

We primarily focus on applying Model 4 (common iron abundance for all AGN model
components, realistic Comptonized continuum) from section \ref{sec_XN1} to all
these observations simultaneously. All parameters that should not vary on
observational timescales, \ie\ the black hole spin, disk inclination and iron abundance,
are linked across all the datasets. In addition to these, we assume that the flux of the
distant reflection is constant across all epochs, and also assume that the primary
continuum parameters ($\Gamma, kT_{\rm{e}}$), and the ionisation states for both the
accretion disk and the ionised absorption are constant across all epochs, given that the
intrinsic flux variations between all the different epochs appear to be quite minor.
Finally, given the strong similarity between the X1, X2 and S1 3--10\,keV spectra, we
also assume that the neutral absorption parameters are the same for these three
epochs.

Within this framework, we test two possible explanations for the AGN variability that
is present in addition to the absorption variability observed. First, we assume that the
geometric structure of the inner accretion flow is static, and that all the variability is
produced through intrinsic changes in the brightness of the corona. In modelling terms,
we therefore keep $h$ constant across all epochs, and allow the normalisation of the
\relxilllpcp\ component to vary. Second, we assume that the intrinsic brightness of the
corona is stable, and that the observed flux variations are produced through geometric
changes that vary the degree of lightbending and thus the fraction of the primary
emission that escapes directly to the observer, \ie\ we allow $h$ to vary and keep the
\relxilllpcp\ normalisation constant. Lower source heights result in stronger lightbending,
reducing the observed flux of the primary continuum (\eg\ \citealt{lightbending}). Of
course, both of these processes can occur simultaneously so the reality likely lies
somewhere in between, but these are useful idealisations that reduce parameter
degeneracies and show the two limiting scenarios. This is broadly similar to the
approach taken in \cite{Walton17v404}.

The global fit obtained for the first scenario (intrinsic variations) is very good
(\chisq/DoF = 1673/1545), and the parameter constraints are given in Table
\ref{tab_param_all}. Many of the results from this analysis are broadly similar to those
obtained with Model 4 considering epoch XN1 alone. As expected, we see significant
variations in the neutral absorption column comparing all the epochs. However, even
during the most absorbed epoch (S2), the best-fit column for the line-of-sight
absorption is not Compton-thick. This is further supported by the fact that there is
reasonable evidence that the ionized absorption -- which must arise closer to the black
hole than the neutral absorber -- is still visible in the data from this epoch; removing
this component from the fit to S2 degrades the fit by $\Delta\chi^{2}$ = 13 for 2 fewer
free parameters.

\begin{figure}
\begin{center}
%\hspace*{-0.5cm}
\rotatebox{0}{
{\includegraphics[width=235pt]{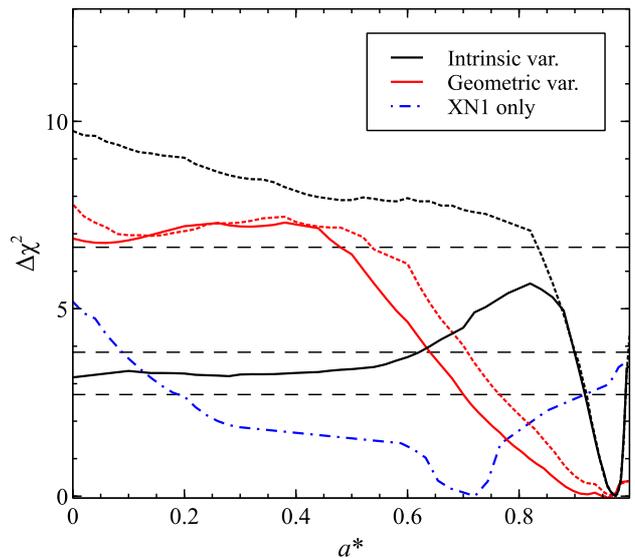}}
}
\end{center}
\caption{
The $\Delta$\chisq\ confidence contours for the black hole spin obtained for our
multi-epoch analysis using Model 4. The black and red (grey in black-and-white)
curves show the results assuming the AGN variability is dominated by intrinsic and
geometric changes in the corona, respectively, and the solid and dotted lines show
the results for a free electron temperature and assuming $kT_{\rm{e}} = 50$\,keV,
respectively. As before, horizontal dashed lines represent the 90, 95 and 99\%
confidence levels for a single parameter of interest. We also show the contour for
epoch XN1 only from Figure \ref{fig_spin_XN1} for comparison (dash-dot blue).}
\label{fig_spin_ME}
\end{figure}

For the black hole spin, we do indeed find that the formal parameter constraints
are tighter than for XN1 alone; a comparison of the confidence contours obtained for
XN1 and our multi-epoch analysis is shown in Figure \ref{fig_spin_ME}. In this case,
the model prefers a high \Rfrac, high spin scenario, with the spin constrained to $a^* >
0.92$. However, as the spin decreases the model quickly enters a region of parameter
space in which the fit is not sensitive to this parameter, and a non-rotating black hole is
again not excluded at high confidence. This is related to the same parameter
degeneracy between \Rfrac\ and $kT_{\rm{e}}$ discussed above (Section
\ref{sec_compt}). At low spins, we find that the best-fit electron temperature runs to
20\,keV, the lower limit of the current \relxilllpcp\ grids. Such low coronal temperatures
are highly unusual among the unobscured AGN population (see e.g. \citealt{Fabian15},
though rare exceptions may be possible, \citealt{Tortosa17, Kara17}). We therefore
re-compute the confidence contour for the spin holding $kT_{\rm{e}}$ fixed at 50\,keV
(also shown in Figure \ref{fig_spin_ME}), which is both the best-fit value for this
scenario, and also roughly the average value observed from the unobscured AGN
compiled in \cite{Fabian15}. In this case, the formal 90\% statistical constraint is
unchanged, but we see that low values for the black hole spin can be more confidently
excluded. We stress that despite the degeneracy between \Rfrac\ and $kT_{\rm{e}}$,
the disk reflection contribution is strongly required by the data. Removing this from the
fit (\ie setting \Rfrac\ = 0) degrades the fit by $\Delta\chi^{2} = 102$ for 4 fewer free
parameters.

%\begin{figure}
%\hspace*{-0.5cm}
%\epsscale{1.05}
%\plotone{./figs/iras13197_ME_spin_contours_cabs.eps}
%\caption{The $\Delta$\chisq\ confidence contours for the black hole spin obtained
%for our multi-epoch analysis using Model 4. The black and red curves show the
%results assuming the AGN variability is dominated by intrinsic and geometric
%changes in the corona, respectively, and the solid and dotted lines show the
%results for a free electron temperature and assuming $kT_{\rm{e}} = 50$\,keV,
%respectively. As before, horizontal dashed lines represent the 90, 95 and 99\%
%confidence levels for a single parameter of interest. We also show the contour
%for epoch XN1 only from Figure \ref{fig_spin_XN1} for comparison (blue).}
%\vspace{0.3cm}
%\label{fig_spin_ME}
%\end{figure}

The second scenario (geometric variations) also provides an equally good fit
(\chisq/DoF = 1667/1545). The parameter constraints are again given in Table
\ref{tab_param_all} and, aside from the enforced differences, there are many
similarities with the intrinsic variations scenario. In particular, the absorption variations
are practically identical between the two scenarios. With this scenario, we find that
the spin is constrained to be $a^* > 0.70$ (see Figure \ref{fig_spin_ME}). This is not
quite as tight as the intrinsic variations scenario, but the degeneracy between \Rfrac\
and $kT_{\rm{e}}$ is not as severe in this case. This is because the best-fit electron
temperature is already unusually low ($\sim$23\,keV), so the model cannot decrease
$kT_{\rm{e}}$ to this level in order to fit lower reflection fractions, and thus lower spins
(we stress though that while $kT_{\rm{e}}$ is already low, as we decrease the spin the
model does not ever hit the lower limit of 20\,keV in the current version of \relxilllpcp,
resulting in artificial increases in \chisq). However, since the best-fit electron
temperature is unusually low, we again re-compute the spin contour with $kT_{\rm{e}}
= 50$\,keV. This worsens the fit a little (\chisq/DoF = 1670/1546) and slightly contracts
the formal spin constraint to $a^* > 0.78$, but otherwise does not result in any other
major changes to the confidence contour.

Finally, we also re-visit the scenario in which the distant components are assumed
to have a solar abundance with the multi-epoch data. Here, we continue to use the 
reflection models that assume realistic Comptonized spectra as the illuminating
continuum, and also the broader approach taken in this section in terms of the
parameters assumed to be constant and to vary between epochs, but return to
decoupling the abundances of the disc components and the distant components,
forcing the latter to have a solar abundance. As with our analysis of epoch XN1, we
again find that the abundance of the disc components becomes highly super-solar
($A_{\rm{Fe}} > 9.0$). However, even if we allow both $h$ and the \relxilllpcp\
normalisation to vary simultaneously, with the multi-epoch data the fit is notably
worse than either of the cases in which all the model components are assumed to
have a common iron abundance: $\chi^{2}$/DoF = 1719/1545, \ie\ $\Delta\chi^{2} >
45$ worse despite having \textit{more} degrees of freedom. We can therefore
confidently exclude this possibility thanks to the multi-epoch dataset, and so we do
not present the results for this model in any further detail.

\section{Discussion}
\label{sec_dis}

We have presented a broadband X-ray spectral analysis of a new coordinated
\xmm+\nustar\ observation of the type 1.8 Seyfert galaxy \iras, along with a
multi-epoch X-ray analysis further incorporating archival observations taken with
\xmm\ and \suzaku. The broadband X-ray spectrum exhibited by this source is
highly complex, with strong contributions from relativistic reflection from the inner
accretion disk, absorption and further reprocessing by more distant material,
and ionised absorption from an outflow all combining to sculpt the observed
spectral form, similar to the well-studied AGN in NGC\,1365 (\citealt{Risaliti13nat,
Walton14, Rivers15}). By combining the high S/N broadband data provided by
\xmm\ and \nustar\ with the multi-epoch archival data obtained with \xmm\ and
\suzaku, we are able to disentangle the relative contributions from all these various
components for a variety of different scenarios, allowing us to place constraints on
the parameters of the innermost accretion flow, and in turn the spin of the black
hole, despite the heavy absorption present in this system.

\subsection{Metallicity and the Inner Disk}

Based on our analysis of epoch XN1, the key inner disk parameters (\eg\ black
hole spin, disk inclination) appear to be largely independent of the precise
treatment of the distant reflection (slab vs torus geometry) and are only moderately
sensitive to the form of the primary continuum (powerlaw with an exponential cutoff
vs realistic Comptonized continuum). However, the results are strongly dependent
on the treatment of the iron abundance for the neutral absorption (and by
extension the distant reflection, which we assume to be associated with the same 
medium).

We test two different scenarios for the iron abundance, focusing initially on epoch
XN1 as the highest S/N broadband dataset. First we assume that the neutral
absorption/distant reprocessor has a solar iron abundance -- as is often the case in
the literature -- but allow the abundance for the disk components to vary, and second
we assume that all the model components have a common iron abundance, which is
free to vary. In the former scenario, we find that the contribution from the disk reflection
needs to be very strong (\Rfrac\ $\gtrsim$ 3) in order to model the strong drop in the
spectrum at $\sim$7\,keV and the curvature of the continuum above 10\,keV (see
Figure \ref{fig_spec}). This requires the X-ray source to be very close to a rapidly
rotating black hole, such that strong gravitational lightbending can sufficiently enhance
the disk reflection relative to the observed primary continuum emission (\eg\ 
\citealt{lightbending}). Combining the constraints from the slab and torus models for
the distant reflection (Models 1 and 2, respectively), we find the black hole spin to be
$0.89 \leq a^* \leq 0.98$ in this scenario. However, in this scenario a large discrepancy
between the iron abundance assumed for the distant material and that inferred for the
disk is seen, with the disk required to be strongly super-solar ($A_{\rm{Fe}} > 7.1$).
Assuming instead that all model components are chemically homogeneous, we find a
moderately super-solar iron abundance ($A_{\rm{Fe}} \sim 2.3$) and that the
requirement for strong reflection and gravitational lightbending is much less severe
(although such solutions are still permitted). As such, the constraints on the spin are
subsequently much looser; considering epoch XN1 only, we could only constrain the
spin to $0.19 \leq a^* \leq 0.98$. Neither of these scenarios is strongly preferred over
the other in a statistical sense for the broadband XN1 data, although the common iron
abundance scenario does result in a minor improvement to the fit (see Table
\ref{tab_param_XN1}).

The scenario in which the iron abundance is common to all AGN model components
is clearly the more intuitive of these two possibilities, and would likely be the
preferred scenario in the situation where only the broadband data from epoch XN1
were available. In addition, the higher inclination inferred in the chemically
homogeneous scenario ($\sim$60\deg) is probably more in line with expectation for
a system with such high levels of absorption. However, the scenario in which the
distant components have a solar abundance cannot statistically be ruled out with
these data alone. Furthermore, although such measurements are notoriously
challenging, the metallicity indicator based on the \ciii, \civ\ and \heii\ lines presented
by \cite{Dors14} suggests that the narrow line region (NLR) in \iras\ has a solar
metallicity. A solar iron abundance could therefore also be expected for the distant
model components in our work, which would still appear to be in some tension with the
super-solar abundance inferred in the chemically homogeneous scenario. There may
therefore be some justification for preferring the chemically inhomogeneous scenario if
both are statistically permissible. This would in turn require a mechanism by which the
inner accretion disk could have, or at least appear to have a significantly different
abundance than the neutral absorber/distant reprocessor.

Two potentially interesting possibilities for such a mechanism are discussed in
\cite{Reynolds12}. The first is that the atomic iron in the distant absorber/reprocessor
could be depleted into dust grains. Should these grains be sufficiently large,
self-shielding could reduce the strength of the atomic iron signatures relative to the
contribution of the distant backscattered continuum (i.e. the distant Compton hump),
which would give the appearance of a reduced iron abundance for the distant model
components. Second, \cite{Reynolds12} also discuss the possibility that the
photospheric iron abundance of the inner disk could appear to be enhanced by
radiative levitation of iron atoms to the disk surface. In this scenario, the background
radiation field within the innermost regions of an AGN accretion disk could produce a
net upwards force on moderately ionised iron, causing the iron to diffuse upwards.
Some evidence for this process may have been observed recently from the Galactic
black hole X-ray binary GRS\,1915+105, where the iron abundance appears to
change across the phase of the 50\,s limit cycle oscillations observed from this source
(\citealt{Zoghbi16}).

These possibilities are not without their problems. With regards to potential dust
depletion of iron in the distant absorber/reprocessor, while the NLR should reside
beyond the dust sublimation radius (\citealt{Netzer15rev}), we note that there is
evidence that the majority of the narrow core to the iron emission in AGN may arise
from regions interior to this (\citealt{Gandhi15}). Furthermore, \cite{Reynolds12} note
that radiative levitation of iron may only be relevant for high accretion rates (close to
the Eddington limit) onto lower mass AGN, where disk temperatures are higher. The
mass of \iras\ is not well known, but \cite{Vasudevan10} estimate the mass and
bolometric luminosity to be $\log (M_{\rm{BH}}/\msun) \sim 7.81$ and $L_{\rm{bol}}
\sim 5-8 \times 10^{44}$\,\ergps, respectively, combining X-ray and infrared
information. These values would correspond to $L_{\rm{bol}}/L_{\rm{Edd}} \sim
0.05-0.1$. While this is still fairly uncertain, the accretion rate is likely too low for
radiative levitation to play an important role in this case. However, although these
processes may struggle individually to produce a metallicity gradient large enough to
match that inferred here, they could potentially act in combination. Furthermore, there
may be other mechanisms beyond those considered here that can produce
strong/additional metallicity gradients (actual or apparent), so it would also be difficult
to conclusively rule out the chemically inhomogeneous scenario through physical
arguments based on epoch XN1 alone.

However, we find that when considering the multi-epoch dataset -- additionally
including archival observations from \xmm\ and \suzaku\ (epochs X1, X2, S1, S2)
which show both varying levels of absorption and intrinsic variability from the central
AGN -- we can distinguish between these two scenarios in a statistical sense, with
the chemically homogeneous scenario clearly preferred. This clearly demonstrates
the importance of combining high S/N broadband observations with a multi-epoch
approach to disentangle the various emission and absorption components for these
complex AGN, particularly when the level of absorption is variable, and the typical
column is as high as it is for \iras.

Within the chemically homogeneous scenario, we test two different possible
explanations for the AGN variability observed. First we assume this is dominated by
intrinsic brightness variations in the corona, and second we assume
it is dominated by variations in the geometry of the corona that change the degree of
lightbending, resulting in variations in the observed flux. Both of these limiting
scenarios fit the multi-epoch data similarly well, and both improve the formal spin
constraint in comparison to the fits to epoch XN1 alone. When the variations are
assumed to be intrinsic, we find $a^* > 0.92$, and when they are assumed to be
geometric we find $a^* > 0.70$. In these fits, we often find that the electron
temperature moves into an unusually low area of parameter space ($kT_{\rm{e}} \sim
20$\,keV). If we make sensible assumptions regarding this temperature based on the
results seen in unobscured systems (\eg\ \citealt{Fabian15}, and references therein)
and assume $kT_{\rm{e}} = 50$\,keV, in the intrinsic variations scenario we find that
slowly rotating black holes can be more confidently excluded (but the formal 90\% spin
constraint is unchanged), and in the geometric variations scenario the spin
constraint improves to $a^* > 0.78$. However, we stress that a slowly rotating black
hole is not very strongly excluded for any of these scenarios, so the spin constraint
presented here would certainly benefit from confirmation with further broadband
observations, ideally probing lower levels of obscuration than seen during epoch XN1.
To be conservative we take $a^* > 0.7$ as our final spin constraint.

\subsection{Variations in the Neutral Absorber}

From our multi-epoch analysis we find evidence that the line-of-sight column of the
neutral absorber varies from epoch to epoch, ranging over $N_{\rm{H}} \sim 5-10
\times 10^{23}$\,cm$^{-2}$, suggesting that the absorbing medium has a clumpy
structure. A hint of this was previously claimed by \cite{Risaliti02} comparing archival
observations by \bepposax\ and \asca, but the subsequent analysis by
\cite{Miniutti07iras} found the column density between these observations and the
first \xmm\ epoch (X1) to be consistent. Although we have made simplifying
assumptions in our analysis, \eg\ forcing the photon index and the height of the X-ray
reflector to be constant across all epochs, the changes in \nh\ observed here are too
strong for these assumptions to have any major influence on this conclusion.
Assuming this absorption is related to the torus structure invoked in AGN unification
schemes, this could add further observational evidence to the suggestion that this
torus is clumpy, rather than a uniform structure (\eg\ \citealt{Nikutta09, Markowitz14}).
However, it is also possible that the absorbing clouds are located instead in the
broad line region, as appears to be the case in NGC\,1365 (\eg\ \citealt{Risaliti09b}).
Unfortunately, the observations presented here are separated by $\sim$years, so a
meaningful test of the location of the absorbing clouds is not currently possible.

If the neutral absorber is associated with the distant reflector, as we have generally
assumed here, then the apparent stability of this emission over the full set of
observations could suggest that this material is located at distances $\gtrsim$10 light
years, or equivalently $\gtrsim$10$^{6}$\,\rg\ for a mass of $\log (M_{\rm{BH}}/\msun)
\sim 7.81$, given the full temporal separation of the observations considered. This
would likely place the material beyond the broad line region (\citealt{Netzer15rev}).
However, we note again that the variations in the intrinsic flux between epochs are
fairly minor (less than a factor of 2, after accounting for the scattering losses; see the
variations in the \relxilllp\ normalisation in Table \ref{tab_param_all}), so it may be that
the average flux on long timescales is relatively constant. If this is the case, then we
may not expect to see strong variations in the distant reflector, even if this material is
located at radii smaller than $10^{6}$\,\rg. All of this could be tested in the future with
higher-cadence broadband observations, both to study the intrinsic flux variations and
the timescales on which the absorption varies.

In contrast to the column density, the covering fraction obtained for the neutral
absorber is persistently high ($\gtrsim$97\%). Covering fractions this high are likely
indicative that, although the model is formally partially covering, along our line-of-sight
the absorber is actually fully covering in a geometric sense, and the non-unity covering
fraction is accounting for the small fraction of the intrinsic continuum emission that gets
scattered around the absorber and back into our line-of-sight. Indeed, X-ray obscured
AGN often display scattered fractions at the $\sim$few percent level (\eg\
\citealt{Ichikawa12}), similar to that inferred here.

\subsection{An Ionised Outflow}

In addition to the relativistic reflection and strong neutral absorption that strongly
influence the broadband spectrum observed, we also find robust evidence that
\iras\ exhibits blueshifted absorption from ionized iron, with absorption lines from
both \fexxv\ and \fexxvi\ observed, confirming the tentative indication reported
in previous works (\citealt{Dadina04, Miniutti07iras}). Evidence for this absorption
is seen at all epochs analysed in this work, but its properties appear to be variable,
with the line-of-sight column density and outflow velocity ranging from $N_{\rm{H}}
\sim 2-6 \times 10^{23}$\,cm$^{-2}$ (assuming the system is chemically
homogeneous) and from $v_{\rm{out}} \sim 5000-8000$\,\kmps, respectively. 
These velocities are not dissimilar to the ionised outflow seen in NGC\,1365
(\citealt{Risaliti05b}), but are not as extreme as the `ultra-fast' outflows now robustly
confirmed with broadband spectroscopy in a few AGN (\eg\ \citealt{Nardini15,
Lobban16, Parker17nat}).

Throughout this work we have assumed that the ionised absorber is associated
with the accretion disc, and therefore physically interior to the neutral absorber. Given
the high degree of ionisation, and the fairly rapid outflow velocity, it is natural to assume
that this absorption would occur closer to the central ionising source. However, it is
worth noting that since we use multiplicative absorption models, and both apply to the
central emission from the source, the results obtained for the inner disc do not depend
on this assumption for any of the models considered. The only issue this relates to is
the iron abundance assumed for the ionised absorption in the chemically
inhomogeneous models (\ie whether this should be linked to the disc components or
the distant components). However, since the only features the ionised absorption
contributes to the observed spectrum are the iron absorption lines, the only result that
will be influenced by this assumption is the column density inferred for this component
(which is strongly degenerate with the iron abundance as a consequence). In the
(preferred) chemically homogenous models the absorbers are fully commutative in a
functional sense, since the iron abundance is the same for all model components.

Although the values obtained for the column density of the ionised absorption do
depend heavily on our treatment of the iron abundance, this only serves to set the
overall scale for the column density results, and cannot explain the variations in the
column seen between epochs. This variability suggests there is also some level of
inhomogeneity in the ionized wind, in addition to the neutral absorption, although the
variations in the ionized absorption do not obviously appear to correlate (or
anti-correlate) with those in the neutral absorber. Such inhomogeneities are likely
also expected for disk winds; MHD simulations of disc winds show them to be
clumpy, time-variable structures (\eg\ \citealt{Proga04}). However, the fact that this
ionised absorption is seen at all epochs likely suggests that along our line-of-sight,
these inhomogeneities may not be too severe.

\section{Conclusions}

The X-ray spectrum of \iras\ is highly complex, exhibiting contributions from relativistic
reflection from the inner accretion disk, absorption and further reprocessing by more
distant material, and absorption from an ionised outflow. By utilizing the high sensitivity,
broadband coverage provided by \xmm+\nustar, as well as a multi-epoch approach
incorporating the archival observations performed by \xmm\ and \suzaku, we perform
detailed spectral analysis with an emphasis on separating out the reflection from the
innermost accretion disk, which previous works have suggested may dominate the
AGN emission. Using the latest reflection and absorption models and focusing first on
the broadband \xmm+\nustar\ epoch, we find that the results for the inner disk do
not strongly depend on the geometry assumed for the distant reprocessor, or the
precise form of the illuminating X-ray continuum. However, these results do depend on
the treatment of the iron abundance of the distant absorber/reprocessor. If this is
assumed to have a solar abundance, as may be suggested by independent metallicity
estimates for the narrow line region, a high-spin, reflection-dominated scenario is
strongly required ($0.89 \leq a^* \leq 0.98$), but the results also require a highly 
super-solar abundance for the disk ($A_{\rm{Fe}} > 7.1$). If we instead assume the
system is chemically homogeneous and link the iron abundance between all the AGN
components, we find a moderately super-solar abundance ($A_{\rm{Fe}} =
2.3^{+0.2}_{-0.1}$) and the constraint on the spin is significantly weaker ($0.19 \leq
a^* \leq 0.98$).

Both of these scenarios fit the \xmm+\nustar\ data similarly well. However, when we
incorporate the data from archival \xmm\ and \suzaku\ observations, and fit these
datasets simultaneously with the \xmm+\nustar\ epoch, we find the chemically
homogeneous scenario is preferred in a statistical sense thanks to the absorption
variability observed. This demonstrates the importance of combining both broadband
and multi-epoch spectroscopy for AGN with sources as complex as \iras. Including this
data also improves the formal spin constraint to $a^* > 0.7$ for this scenario, so a
rapidly rotating black hole is preferred. However, a slowly rotating black hole is still not
strongly excluded, so this constraint should be confirmed with additional broadband
observations.

In addition to the results for the inner disk, through our multi-epoch analysis we also
find that both the neutral and ionised absorbers vary from epoch to epoch. This
suggests that both the neutral absorber, either the torus or broad line region clouds,
and the ionised absorber, an accretion disk wind, have an inhomogeneous, clumpy
structure (at least to some extent). Higher cadence broadband monitoring in the future
should be able to constrain the location of the variable neutral absorber and determine
whether this is associated with the torus or the broad line region.

\section*{ACKNOWLEDGEMENTS}

The authors would like to thank the reviewer for their helpful feedback, which helped
to improve the final version of the manuscript. DJW acknowledges support from an
STFC Ernest Rutherford Fellowship, and ACF acknowledges support from ERC
Advanced Grant 340442. GM thanks the European Union Seventh Framework
Program (FP7/2007--2013) for funding under grant 312789 (StrongGravity). This
research has made use of data obtained with \nustar, a project led by Caltech,
funded by NASA and managed by NASA/JPL, and has utilized the \nustardas\
software  package, jointly developed by the ASDC (Italy) and Caltech (USA). This
research has also made use of data obtained with \xmm, an ESA science mission
with instruments and contributions directly funded by ESA Member States, and with
\suzaku, a collaborative mission between the space agencies of Japan (JAXA) and
the USA (NASA).

%{\it Facilities:} \facility{NuSTAR}, \facility{XMM-Newton}, \facility{Suzaku}

\bibliographystyle{/Users/dwalton/papers/mnras}

\bibliography{/Users/dwalton/papers/references}

\begin{thebibliography}{85}
\expandafter\ifx\csname natexlab\endcsname\relax\def\natexlab#1{#1}\fi

\bibitem[{Anders} \& {Grevesse}(1989)]{Anders89}
{Anders} E., {Grevesse} N., 1989, \gca, 53, 197

\bibitem[{Arnaud}(1996)]{xspec}
{Arnaud} K.~A., 1996, in { Astronomical Data Analysis Software and Systems
  V\/}, edited by {G.~H.~Jacoby \& J.~Barnes}, vol. 101 of { Astron. Soc. Pac.
  Conference Series, Astron. Soc. Pac., San Francisco\/}, ~17

\bibitem[{Boldt}(1987)]{Boldt87}
{Boldt} E., 1987, in { Observational Cosmology\/}, edited by {A.~Hewitt,
  G.~Burbidge, \& L.~Z.~Fang}, vol. 124 of { IAU Symposium\/},  611--615

\bibitem[{Bonson} \& {Gallo}(2016)]{Bonson16}
{Bonson} K., {Gallo} L.~C., 2016, \mnras, 458, 1927

\bibitem[{Brenneman} et~al.(2011){Brenneman}, {Reynolds}, {Nowak}
  et~al.]{Brenneman11}
{Brenneman} L.~W., {Reynolds} C.~S., {Nowak} M.~A., et~al., 2011, \apj, 736,
  103

\bibitem[{Brightman} et~al.(2015){Brightman}, {Balokovi{\'c}}, {Stern}
  et~al.]{Brightman15}
{Brightman} M., {Balokovi{\'c}} M., {Stern} D., et~al., 2015, \apj, 805, 41

\bibitem[{Brightman} \& {Nandra}(2011)]{torus}
{Brightman} M., {Nandra} K., 2011, \mnras, 413, 1206

\bibitem[{Dadina} \& {Cappi}(2004)]{Dadina04}
{Dadina} M., {Cappi} M., 2004, \aap, 413, 921

\bibitem[{Dauser} et~al.(2014){Dauser}, {Garc{\'{\i}}a}, {Parker}, {Fabian} \&
  {Wilms}]{Dauser14}
{Dauser} T., {Garc{\'{\i}}a} J., {Parker} M.~L., {Fabian} A.~C., {Wilms} J.,
  2014, \mnras, 444, L100

\bibitem[{Dauser} et~al.(2016){Dauser}, {Garc{\'{\i}}a}, {Walton}
  et~al.]{relxill_norm}
{Dauser} T., {Garc{\'{\i}}a} J., {Walton} D.~J., et~al., 2016, \aap, 590, A76

\bibitem[{Dauser} et~al.(2010){Dauser}, {Wilms}, {Reynolds} \&
  {Brenneman}]{relconv}
{Dauser} T., {Wilms} J., {Reynolds} C.~S., {Brenneman} L.~W., 2010, \mnras,
  409, 1534

\bibitem[{Dors} et~al.(2014){Dors}, {Cardaci}, {H{\"a}gele} \&
  {Krabbe}]{Dors14}
{Dors} O.~L., {Cardaci} M.~V., {H{\"a}gele} G.~F., {Krabbe} {\^A}.~C., 2014,
  \mnras, 443, 1291

\bibitem[{Dubois} et~al.(2014){Dubois}, {Volonteri} \& {Silk}]{Dubois14}
{Dubois} Y., {Volonteri} M., {Silk} J., 2014, \mnras, 440, 1590

\bibitem[{Fabian} et~al.(2013){Fabian}, {Kara}, {Walton} et~al.]{Fabian13iras}
{Fabian} A.~C., {Kara} E., {Walton} D.~J., et~al., 2013, \mnras, 429, 2917

\bibitem[{Fabian} et~al.(2015){Fabian}, {Lohfink}, {Kara}, {Parker},
  {Vasudevan} \& {Reynolds}]{Fabian15}
{Fabian} A.~C., {Lohfink} A., {Kara} E., {Parker} M.~L., {Vasudevan} R.,
  {Reynolds} C.~S., 2015, \mnras, 451, 4375

\bibitem[{Fabian} et~al.(1989){Fabian}, {Rees}, {Stella} \& {White}]{Fabian89}
{Fabian} A.~C., {Rees} M.~J., {Stella} L., {White} N.~E., 1989, \mnras, 238,
  729

\bibitem[{Fabian} et~al.(2009){Fabian}, {Zoghbi}, {Ross} et~al.]{FabZog09}
{Fabian} A.~C., {Zoghbi} A., {Ross} R.~R., et~al., 2009, \nat, 459, 540

\bibitem[{F{\"u}rst} et~al.(2016){F{\"u}rst}, {M{\"u}ller}, {Madsen}
  et~al.]{Fuerst16}
{F{\"u}rst} F., {M{\"u}ller} C., {Madsen} K.~K., et~al., 2016, \apj, 819, 150

\bibitem[{Gandhi} et~al.(2015){Gandhi}, {H{\"o}nig} \& {Kishimoto}]{Gandhi15}
{Gandhi} P., {H{\"o}nig} S.~F., {Kishimoto} M., 2015, \apj, 812, 113

\bibitem[{Garc{\'{\i}}a} et~al.(2014){Garc{\'{\i}}a}, {Dauser}, {Lohfink}
  et~al.]{relxill}
{Garc{\'{\i}}a} J., {Dauser} T., {Lohfink} A., et~al., 2014, \apj, 782, 76

\bibitem[{Garc{\'{\i}}a} \& {Kallman}(2010)]{xillver}
{Garc{\'{\i}}a} J., {Kallman} T.~R., 2010, \apj, 718, 695

\bibitem[{Garc{\'{\i}}a} et~al.(2015){Garc{\'{\i}}a}, {Dauser}, {Steiner},
  {McClintock}, {Keck} \& {Wilms}]{Garcia15}
{Garc{\'{\i}}a} J.~A., {Dauser} T., {Steiner} J.~F., {McClintock} J.~E., {Keck}
  M.~L., {Wilms} J., 2015, \apjl, 808, L37

\bibitem[{Grevesse} \& {Sauval}(1998)]{Grevesse98}
{Grevesse} N., {Sauval} A.~J., 1998, \ssr, 85, 161

\bibitem[{Haardt} \& {Maraschi}(1991)]{Haardt91}
{Haardt} F., {Maraschi} L., 1991, \apjl, 380, L51

\bibitem[{Harrison} et~al.(2013){Harrison}, {Craig}, {Christensen}
  et~al.]{NUSTAR}
{Harrison} F.~A., {Craig} W.~W., {Christensen} F.~E., et~al., 2013, \apj, 770,
  103

\bibitem[{Ichikawa} et~al.(2012){Ichikawa}, {Ueda}, {Terashima}
  et~al.]{Ichikawa12}
{Ichikawa} K., {Ueda} Y., {Terashima} Y., et~al., 2012, \apj, 754, 45

\bibitem[{Jansen} et~al.(2001){Jansen}, {Lumb}, {Altieri} et~al.]{XMM}
{Jansen} F., {Lumb} D., {Altieri} B., et~al., 2001, \aap, 365, L1

\bibitem[{Kalberla} et~al.(2005){Kalberla}, {Burton}, {Hartmann} et~al.]{NH}
{Kalberla} P.~M.~W., {Burton} W.~B., {Hartmann} D., et~al., 2005, \aap, 440,
  775

\bibitem[{Kallman} \& {Bautista}(2001)]{xstar}
{Kallman} T., {Bautista} M., 2001, \apjs, 133, 221

\bibitem[{Kara} et~al.(2017){Kara}, {Garcia}, {Lohfink} et~al.]{Kara17}
{Kara} E., {Garcia} J.~A., {Lohfink} A., et~al., 2017, ArXiv e-prints

\bibitem[{Koyama} et~al.(2007){Koyama}, {Tsunemi}, {Dotani} et~al.]{SUZAKU_XIS}
{Koyama} K., {Tsunemi} H., {Dotani} T., et~al., 2007, \pasj, 59, 23

\bibitem[{Lansbury} et~al.(2017){Lansbury}, {Alexander}, {Aird}
  et~al.]{Lansbury17}
{Lansbury} G.~B., {Alexander} D.~M., {Aird} J., et~al., 2017, ArXiv 1707.06651

\bibitem[{Laor}(1991)]{kdblur}
{Laor} A., 1991, \apj, 376, 90

\bibitem[{Lobban} et~al.(2016){Lobban}, {Pounds}, {Vaughan} \&
  {Reeves}]{Lobban16}
{Lobban} A.~P., {Pounds} K., {Vaughan} S., {Reeves} J.~N., 2016, \apj, 831, 201

\bibitem[{Madsen} et~al.(2015){Madsen}, {Harrison}, {Markwardt}
  et~al.]{NUSTARcal}
{Madsen} K.~K., {Harrison} F.~A., {Markwardt} C.~B., et~al., 2015, \apjs, 220,
  8

\bibitem[{Mantovani} et~al.(2016){Mantovani}, {Nandra} \& {Ponti}]{Mantovani16}
{Mantovani} G., {Nandra} K., {Ponti} G., 2016, \mnras, 458, 4198

\bibitem[{Markowitz} et~al.(2014){Markowitz}, {Krumpe} \&
  {Nikutta}]{Markowitz14}
{Markowitz} A.~G., {Krumpe} M., {Nikutta} R., 2014, \mnras, 439, 1403

\bibitem[{Miller} et~al.(2008){Miller}, {Turner} \& {Reeves}]{Miller08}
{Miller} L., {Turner} T.~J., {Reeves} J.~N., 2008, \aap, 483, 437

\bibitem[{Miller} et~al.(2009){Miller}, {Turner} \& {Reeves}]{LMiller09}
{Miller} L., {Turner} T.~J., {Reeves} J.~N., 2009, \mnras, 399, L69

\bibitem[{Miniutti} \& {Fabian}(2004)]{lightbending}
{Miniutti} G., {Fabian} A.~C., 2004, \mnras, 349, 1435

\bibitem[{Miniutti} et~al.(2007){Miniutti}, {Ponti}, {Dadina}, {Cappi} \&
  {Malaguti}]{Miniutti07iras}
{Miniutti} G., {Ponti} G., {Dadina} M., {Cappi} M., {Malaguti} G., 2007,
  \mnras, 375, 227

\bibitem[{Murphy} \& {Yaqoob}(2009)]{mytorus}
{Murphy} K.~D., {Yaqoob} T., 2009, \mnras, 397, 1549

\bibitem[{Nardini} et~al.(2011){Nardini}, {Fabian}, {Reis} \&
  {Walton}]{Nardini11}
{Nardini} E., {Fabian} A.~C., {Reis} R.~C., {Walton} D.~J., 2011, \mnras, 410,
  1251

\bibitem[{Nardini} et~al.(2015){Nardini}, {Reeves}, {Gofford}
  et~al.]{Nardini15}
{Nardini} E., {Reeves} J.~N., {Gofford} J., et~al., 2015, Science, 347, 860

\bibitem[{Netzer}(2015)]{Netzer15rev}
{Netzer} H., 2015, \araa, 53, 365

\bibitem[{Nikutta} et~al.(2009){Nikutta}, {Elitzur} \& {Lacy}]{Nikutta09}
{Nikutta} R., {Elitzur} M., {Lacy} M., 2009, \apj, 707, 1550

\bibitem[{Parker} et~al.(2017){Parker}, {Pinto}, {Fabian} et~al.]{Parker17nat}
{Parker} M.~L., {Pinto} C., {Fabian} A.~C., et~al., 2017, \nat, 543, 83

\bibitem[{Parker} et~al.(2014){Parker}, {Wilkins}, {Fabian}
  et~al.]{Parker14mrk}
{Parker} M.~L., {Wilkins} D.~R., {Fabian} A.~C., et~al., 2014, \mnras, 443,
  1723

\bibitem[{Proga} \& {Kallman}(2004)]{Proga04}
{Proga} D., {Kallman} T.~R., 2004, \apj, 616, 688

\bibitem[{Reis} et~al.(2014){Reis}, {Reynolds}, {Miller} \&
  {Walton}]{Reis14nat}
{Reis} R.~C., {Reynolds} M.~T., {Miller} J.~M., {Walton} D.~J., 2014, \nat,
  507, 207

\bibitem[{Reynolds}(2014)]{Reynolds14rev}
{Reynolds} C.~S., 2014, \ssr, 183, 277

\bibitem[{Reynolds} et~al.(2012){Reynolds}, {Brenneman}, {Lohfink}
  et~al.]{Reynolds12}
{Reynolds} C.~S., {Brenneman} L.~W., {Lohfink} A.~M., et~al., 2012, \apj, 755,
  88

\bibitem[{Reynolds} et~al.(2014){Reynolds}, {Walton}, {Miller} \&
  {Reis}]{Reynolds14}
{Reynolds} M.~T., {Walton} D.~J., {Miller} J.~M., {Reis} R.~C., 2014, \apjl,
  792, L19

\bibitem[{Risaliti} et~al.(2005){Risaliti}, {Bianchi}, {Matt}
  et~al.]{Risaliti05b}
{Risaliti} G., {Bianchi} S., {Matt} G., et~al., 2005, \apjl, 630, L129

\bibitem[{Risaliti} et~al.(2002){Risaliti}, {Elvis} \& {Nicastro}]{Risaliti02}
{Risaliti} G., {Elvis} M., {Nicastro} F., 2002, \apj, 571, 234

\bibitem[{Risaliti} et~al.(2013){Risaliti}, {Harrison}, {Madsen}
  et~al.]{Risaliti13nat}
{Risaliti} G., {Harrison} F.~A., {Madsen} K.~K., et~al., 2013, \nat, 494, 449

\bibitem[{Risaliti} et~al.(2009){Risaliti}, {Salvati}, {Elvis}
  et~al.]{Risaliti09b}
{Risaliti} G., {Salvati} M., {Elvis} M., et~al., 2009, \mnras, 393, L1

\bibitem[{Rivers} et~al.(2013){Rivers}, {Markowitz} \& {Rothschild}]{Rivers13}
{Rivers} E., {Markowitz} A., {Rothschild} R., 2013, \apj, 772, 114

\bibitem[{Rivers} et~al.(2015){Rivers}, {Risaliti}, {Walton} et~al.]{Rivers15}
{Rivers} E., {Risaliti} G., {Walton} D.~J., et~al., 2015, \apj, 804, 107

\bibitem[{Sesana} et~al.(2014){Sesana}, {Barausse}, {Dotti} \&
  {Rossi}]{Sesana14}
{Sesana} A., {Barausse} E., {Dotti} M., {Rossi} E.~M., 2014, ArXiv e-prints

\bibitem[{Sim} et~al.(2010){Sim}, {Proga}, {Miller}, {Long} \& {Turner}]{Sim10}
{Sim} S.~A., {Proga} D., {Miller} L., {Long} K.~S., {Turner} T.~J., 2010,
  \mnras, 408, 1396

\bibitem[{Str{\"u}der} et~al.(2001){Str{\"u}der}, {Briel}, {Dennerl}
  et~al.]{XMM_PN}
{Str{\"u}der} L., {Briel} U., {Dennerl} K., et~al., 2001, \aap, 365, L18

\bibitem[{Takahashi} et~al.(2007){Takahashi}, {Abe}, {Endo} et~al.]{SUZAKU_HXD}
{Takahashi} T., {Abe} K., {Endo} M., et~al., 2007, \pasj, 59, 35

\bibitem[{Tanaka} et~al.(1995){Tanaka}, {Nandra}, {Fabian} et~al.]{Tanaka95}
{Tanaka} Y., {Nandra} K., {Fabian} A.~C., et~al., 1995, \nat, 375, 659

\bibitem[{Tatum} et~al.(2013){Tatum}, {Turner}, {Miller} \& {Reeves}]{Tatum13}
{Tatum} M.~M., {Turner} T.~J., {Miller} L., {Reeves} J.~N., 2013, \apj, 762, 80

\bibitem[{Titarchuk}(1994)]{comptt}
{Titarchuk} L., 1994, \apj, 434, 570

\bibitem[{Tortosa} et~al.(2016){Tortosa}, {Marinucci}, {Matt}
  et~al.]{Tortosa17}
{Tortosa} A., {Marinucci} A., {Matt} G., et~al., 2016, ArXiv 1612.05871

\bibitem[{Turner} et~al.(2001){Turner}, {Abbey}, {Arnaud} et~al.]{XMM_MOS}
{Turner} M.~J.~L., {Abbey} A., {Arnaud} M., et~al., 2001, \aap, 365, L27

\bibitem[{Vasudevan} et~al.(2010){Vasudevan}, {Fabian}, {Gandhi}, {Winter} \&
  {Mushotzky}]{Vasudevan10}
{Vasudevan} R.~V., {Fabian} A.~C., {Gandhi} P., {Winter} L.~M., {Mushotzky}
  R.~F., 2010, \mnras, 402, 1081

\bibitem[{Vasudevan} et~al.(2016){Vasudevan}, {Fabian}, {Reynolds}, {Aird},
  {Dauser} \& {Gallo}]{Vasudevan16}
{Vasudevan} R.~V., {Fabian} A.~C., {Reynolds} C.~S., {Aird} J., {Dauser} T.,
  {Gallo} L.~C., 2016, \mnras, 458, 2012

\bibitem[{Verner} et~al.(1996){Verner}, {Ferland}, {Korista} \&
  {Yakovlev}]{Verner96}
{Verner} D.~A., {Ferland} G.~J., {Korista} K.~T., {Yakovlev} D.~G., 1996, \apj,
  465, 487

\bibitem[{Volonteri} et~al.(2013){Volonteri}, {Sikora}, {Lasota} \&
  {Merloni}]{Volonteri13}
{Volonteri} M., {Sikora} M., {Lasota} J.-P., {Merloni} A., 2013, \apj, 775, 94

\bibitem[{Walton} et~al.(2017){Walton}, {Mooley}, {King} et~al.]{Walton17v404}
{Walton} D.~J., {Mooley} K., {King} A.~L., et~al., 2017, \apj, 839, 110

\bibitem[{Walton} et~al.(2013){Walton}, {Nardini}, {Fabian}, {Gallo} \&
  {Reis}]{Walton13spin}
{Walton} D.~J., {Nardini} E., {Fabian} A.~C., {Gallo} L.~C., {Reis} R.~C.,
  2013, \mnras, 428, 2901

\bibitem[{Walton} et~al.(2010){Walton}, {Reis} \& {Fabian}]{Walton10Hex}
{Walton} D.~J., {Reis} R.~C., {Fabian} A.~C., 2010, \mnras, 408, 601

\bibitem[{Walton} et~al.(2015){Walton}, {Reynolds}, {Miller}, {Reis}, {Stern}
  \& {Harrison}]{Walton15lqso}
{Walton} D.~J., {Reynolds} M.~T., {Miller} J.~M., {Reis} R.~C., {Stern} D.,
  {Harrison} F.~A., 2015, \apj, 805, 161

\bibitem[{Walton} et~al.(2014){Walton}, {Risaliti}, {Harrison}
  et~al.]{Walton14}
{Walton} D.~J., {Risaliti} G., {Harrison} F.~A., et~al., 2014, \apj, 788, 76

\bibitem[{Walton} et~al.(2016){Walton}, {Tomsick}, {Madsen}
  et~al.]{Walton16cyg}
{Walton} D.~J., {Tomsick} J.~A., {Madsen} K.~K., et~al., 2016, \apj, 826, 87

\bibitem[{Wilkins} \& {Fabian}(2012)]{Wilkins12}
{Wilkins} D.~R., {Fabian} A.~C., 2012, \mnras, 424, 1284

\bibitem[{Wilms} et~al.(2000){Wilms}, {Allen} \& {McCray}]{tbabs}
{Wilms} J., {Allen} A., {McCray} R., 2000, \apj, 542, 914

\bibitem[{Xu} et~al.(2017){Xu}, {Balokovic}, {Walton}, {Harrison}, {Garcia} \&
  {Koss}]{Xu17iras}
{Xu} Y., {Balokovic} M., {Walton} D.~J., {Harrison} F.~A., {Garcia} J.~A.,
  {Koss} M.~J., 2017, ArXiv 1702.00073

\bibitem[{Zdziarski} et~al.(1996){Zdziarski}, {Johnson} \&
  {Magdziarz}]{nthcomp1}
{Zdziarski} A.~A., {Johnson} W.~N., {Magdziarz} P., 1996, \mnras, 283, 193

\bibitem[{Zdziarski} et~al.(2003){Zdziarski}, {Lubi{\'n}ski}, {Gilfanov} \&
  {Revnivtsev}]{Zdziarski03}
{Zdziarski} A.~A., {Lubi{\'n}ski} P., {Gilfanov} M., {Revnivtsev} M., 2003,
  \mnras, 342, 355

\bibitem[{Zoghbi} et~al.(2016){Zoghbi}, {Miller}, {King} et~al.]{Zoghbi16}
{Zoghbi} A., {Miller} J.~M., {King} A.~L., et~al., 2016, \apj, 833, 165

\bibitem[{Zycki} et~al.(1999){Zycki}, {Done} \& {Smith}]{nthcomp2}
{Zycki} P.~T., {Done} C., {Smith} D.~A., 1999, \mnras, 309, 561

\end{thebibliography}

\label{lastpage}

\end{document}